\pdfoutput=1
\documentclass[sigconf]{acmart}

\usepackage{subfigure}
\usepackage{multirow}
\usepackage{multicol}
\usepackage[linesnumbered,boxed,ruled,commentsnumbered]{algorithm2e}
\usepackage{enumitem}
\setlist[itemize]{leftmargin=0.4cm}
\usepackage{pifont}
\usepackage{balance} 
\usepackage{xcolor}
\usepackage{color, colortbl}
\definecolor{crimson1}{RGB}{220, 20, 60}

\definecolor{yymgray1}{HTML}{DDDFE8}
\definecolor{yymgray2}{HTML}{F2F2F2}
\definecolor{yympurple}{HTML}{AA93B4}
\definecolor{yymblue}{HTML}{E5F2FC}
\definecolor{yymblue1}{rgb}{0.6, 0.85, 0.95}
\definecolor{yymorange}{HTML}{FDECEE}
\definecolor{yymgreen}{HTML}{e6faef}
\definecolor{yymgreen1}{rgb}{0.7, 0.85, 0.85}
\copyrightyear{2025}
\acmYear{2025}
\setcopyright{acmlicensed}\acmConference[KDD '25]{Proceedings of the 31st ACM SIGKDD Conference on Knowledge Discovery and Data Mining V.2}{August 3--7, 2025}{Toronto, ON, Canada}
\acmBooktitle{Proceedings of the 31st ACM SIGKDD Conference on Knowledge Discovery and Data Mining V.2 (KDD '25), August 3--7, 2025, Toronto, ON, Canada}
\acmDOI{10.1145/3711896.3736936}
\acmISBN{979-8-4007-1454-2/2025/08}

\begin{document}

\title{Enhancing Homophily-Heterophily Separation: Relation-Aware Learning in Heterogeneous Graphs}

\author{Ziyu Zheng}
\orcid{https://orcid.org/0009-0000-3662-0832}
\affiliation{%
  \department{School of Computer Science and Technology,}
  \institution{Xidian University,}
  \city{Xi'an}
  \country{China}
}
\email{zhengziyu@stu.xidian.edu.cn}

\author{Yaming Yang}
\orcid{https://orcid.org/0000-0002-8186-0648}
\affiliation{%
  \department{School of Computer Science and Technology,}
  \institution{Xidian University,}
  \city{Xi'an}
  \country{China}
}
\email{yym@xidian.edu.cn}

\author{Ziyu Guan}
% \authornote{Corresponding Author}
\orcid{https://orcid.org/0000-0003-2413-4698}
\affiliation{%
  \department{School of Computer Science and Technology,}
  \institution{Xidian University,}
  \city{Xi'an}
  \country{China}
}
\email{zyguan@xidian.edu.cn}

\author{Wei Zhao}
\authornote{Corresponding Author}
\orcid{https://orcid.org/0000-0002-9767-1323}
\affiliation{%
  \department{School of Computer Science and Technology,}
  \institution{Xidian University,}
  \city{Xi'an}
  \country{China}
}
\email{ywzhao@mail.xidian.edu.cn}

\author{Weigang Lu}
\orcid{https://orcid.org/0000-0003-4855-7070}
\affiliation{%
  \department{School of Computer Science and Technology,}
  \institution{Xidian University,}
  \city{Xi'an}
  \country{China}
}
\email{wglu@stu.xidian.edu.cn}
\renewcommand{\shortauthors}{Ziyu Zheng, Yaming Yang, Ziyu Guan, Wei Zhao, \& Weigang Lu}
%% No italics

\begin{abstract}
Real-world networks usually have a property of node heterophily, that is, the connected nodes usually have different features or different labels. This heterophily issue has been extensively studied in homogeneous graphs but remains under-explored in heterogeneous graphs, where there are multiple types of nodes and edges. Capturing node heterophily in heterogeneous graphs is very challenging since both node/edge heterogeneity and node heterophily should be carefully taken into consideration. Existing methods typically convert heterogeneous graphs into homogeneous ones to learn node heterophily, which will inevitably lose the potential heterophily conveyed by heterogeneous relations. To bridge this gap, we propose Relation-Aware Separation of Homophily and Heterophily (RASH), a novel contrastive learning framework that explicitly models high-order semantics of heterogeneous interactions and adaptively separates homophilic and heterophilic patterns. Particularly, RASH introduces dual heterogeneous hypergraphs to encode multi-relational bipartite subgraphs and dynamically constructs homophilic graphs and heterophilic graphs based on relation importance. A multi-relation contrastive loss is designed to align heterogeneous and homophilic/heterophilic views by maximizing mutual information. In this way, RASH simultaneously resolves the challenges of heterogeneity and heterophily in heterogeneous graphs. Extensive experiments on benchmark datasets demonstrate the effectiveness of RASH across various downstream tasks. The code is available at: \url{https://github.com/zhengziyu77/RASH}.
\end{abstract}

% \begin{CCSXML}
% <ccs2012>
% <concept>
% <concept_id>10002951.10003227.10003351</concept_id>
% <concept_desc>Information systems~Data mining</concept_desc>
% <concept_significance>500</concept_significance>
% </concept>
% <concept>
% <concept_id>10010147.10010257.10010293.10010294</concept_id>
% <concept_desc>Computing methodologies~Neural networks</concept_desc>
% <concept_significance>500</concept_significance>
% </concept>
% </ccs2012>
% \end{CCSXML}

% \ccsdesc[500]{Computing methodologies~Machine learning}
% \ccsdesc[500]{Networks~Network algorithms}
% %\ccsdesc[500]{Mathematics of computing~Hypergraphs}



\maketitle
\newcommand\kddavailabilityurl{https://doi.org/10.5281/zenodo.15510091}

\ifdefempty{\kddavailabilityurl}{}{
\begingroup\small\noindent\raggedright\textbf{KDD Availability Link:}\\
% please change the following context to include multiple artifacts if necessary.
The source code of this paper has been made publicly available at \url{\kddavailabilityurl}.
\endgroup
}

\section{Introduction}
Heterogeneous graphs, which consist of multiple types of nodes and edges, exhibit more complex semantics compared to homogeneous graphs, and are prevalent in real-world applications such as academic networks ~\cite{han,oag} and social networks~\cite{pre,Trustrecom}. These real-world graphs typically exhibit two key properties: heterogeneity (diverse node/edge types) and heterophily (connections between dissimilar nodes within the same type)~\cite{magnn,heterophily}. Heterogeneous Graph Neural Networks (HGNN)~\cite{hgnn,ie-hgcn} have been shown to effectively capture the structural and semantic patterns of heterogeneous graphs by processing multi-type features at the node level, and are proven effective in addressing heterogeneity. While heterophily has been extensively studied in homogeneous graphs~\cite{fagcn,ACM-GCN,glognn}, the diversity of node and edge types in heterogeneous graphs presents a significant challenge for extracting heterophily in such graphs. This dual challenge underscores the necessity of a unified framework to jointly optimize both heterogeneity and heterophily.

\begin{figure}[ht]
\centering   
\includegraphics[width=1.0\linewidth]{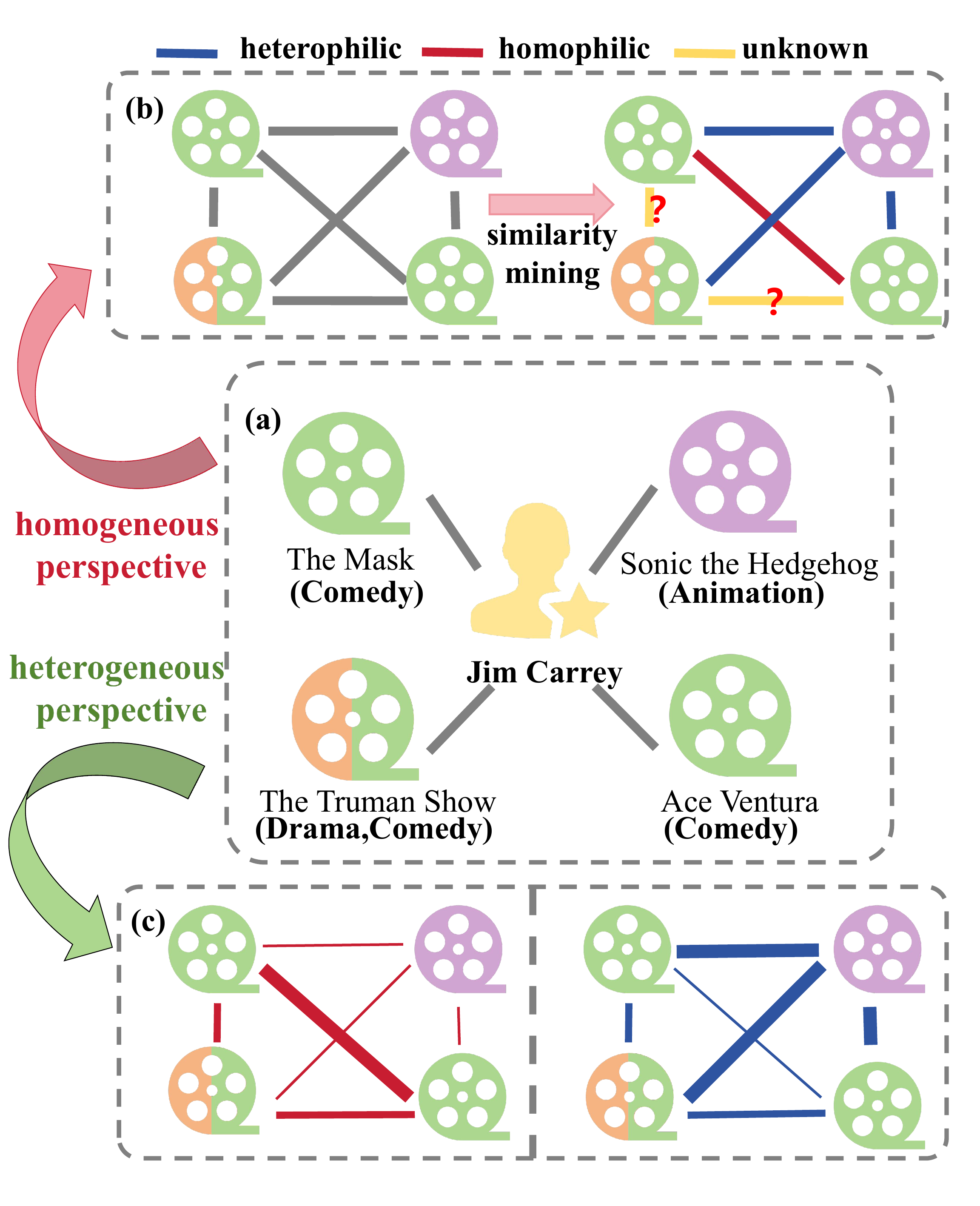} 
\caption{This image is a real movie-actor heterogeneous image, Jim Carrey is a famous comedian, he appeared in the film genre Comedy-based, but also contains other types of movies, such as The Truman Show to the drama-based and contains comedy elements. Even though Sonic the Hedgehog is an animated movie, Jim Carrey plays a comedic role in it. Therefore, we can find a tendency for the movies in which he has starred to be strongly or weakly related to the element of comedy. This relation's importance is vital information for identifying the different classes of movie nodes.} 
\label{Toy.model}
\end{figure}

Although both heterogeneity and heterophily coexist in real-world networks, existing methods predominantly address these properties in isolation. Methods based on HGNNs ~\cite{han,metapath2vec,magnn,rgcn,ie-hgcn} do not consider heterophily. In contrast, recent attempts to model heterophily in heterogeneous graphs adopt a homogeneous perspective via meta-paths ~\cite{latgrl,hdhgr,hetero2net}. These methods rely on feature similarity ~\cite{knn} among nodes of the same type to extract homogeneous graphs for specific types of nodes, learning homophilic and heterophilic patterns from a homogeneous graph perspective. However, this approach overlooks the structure heterogeneity and potential heterophily driven by heterogeneous relations. As shown in Figure ~\ref{Toy.model}(a), in a real-world actor-movie heterogeneous graph, existing approaches tend to use meta-paths or sampling to transform heterogeneous graphs into homogeneous graphs, and then distinguish homophilic edges and heterophilic edges through similarity mining (Figure ~\ref{Toy.model}(b)), It is difficult to distinguish mixed-genre movies relying on the feature similarity of the movie-genre nodes only: The Truman Show, which contains both the drama and comedy genres. It contains both homophilic (Comedy) and heterophilic (Drama) information for its neighbors. This leads to a key question: \textit{can we separate the potential homophily and heterophily of specific types of nodes in heterogeneous graphs without discarding heterogeneity?}

To address the above problem, we propose the Relation-Aware Separation of Homophily and Heterophily (RASH), a contrastive learning framework that dynamically separates homophilic and heterophilic patterns through relation importance. This approach eliminates the reliance on modeling heterophily from a homogeneous perspective. Specifically, we proceed from a heterogeneous perspective, leveraging the importance of heterogeneous edges to learn heterophily. To better capture the structural information of heterogeneous edges, we introduce a dual heterogeneous hypergraph and a hypergraph neural network~\cite{hypergnn} to encode multi-relational bipartite subgraphs, thereby explicitly modeling the higher-order semantics of heterogeneous edges. Further, we dynamically construct homophilic and heterophilic graphs using relation importance scores. For example, in an actor-movie heterogeneous graph, actors tend to prefer certain types of movies. This implies that the weight of an actor's relationship with different types of movies should vary, reflecting their different contributions to movie classification. Based on relation importance, we can assess which movies are more likely to belong to the same category and which are likely to belong to different categories, thus achieving dynamic separation of homophily and heterophily for specific types of nodes in a heterogeneous graph, which is illustrated in Figure ~\ref{Toy.model}(c), we use lines of different thicknesses to represent the weights of homophilic and heterophilic edges. Finally, we introduce a multi-relation contrastive loss to maximize the mutual information between heterogeneous views and homophilic/heterophilic views under different relational guidance, injecting homophilic and heterophilic information into the representations while preserving heterogeneity. This enriches the representations with additional task-relevant information for downstream tasks.

The main contributions of this work are summarized as follows:
\begin{itemize}
\item We formally introduce the joint learning of heterogeneity and heterophily in heterogeneous graphs. The co-existence of semantic heterogeneity, structural heterogeneity, and heterophily in the real world is addressed for the first time. This is distinguished from existing work that does not comprehensively consider these different properties.
\item We introduce a dual heterogeneous hypergraph to explicitly model the complex higher-order semantics of heterogeneous edges. We propose relation-aware importance to dynamically separate homophilic and heterophilic patterns, avoiding semantic loss caused by homogeneous simplifications.
\item We conduct extensive experiments on various benchmark datasets, demonstrating the effectiveness of the proposed method. Our approach shows superior performance across multiple downstream tasks and proves the robustness of RASH, even in the absence of real feature information.
\end{itemize}

\section{Related Work}
\subsection{Heterogeneous Graph Self-Supervised Learning}
Heterogeneous graph self-supervised learning can be classified into metapath-based and free-metapath methods. Meta-path-based methods transform into homogeneous graphs by predefined metapaths. DMGI ~\cite{dmgi} and HDMI ~\cite{hdmi} learn node consistency by maximizing node-level and graph representations' mutual information. HeCo ~\cite{heco} and MEOW ~\cite{meow} introduce network schema perspective and metapath perspective for contrastive learning. HGMAE ~\cite{hgmae} reconstructs the metapath-based edges, and node features from a generative perspective. These methods rely on expert knowledge of predefined metapaths, introducing expensive costs. 

In the meta-path-free method, SHGP ~\cite{shgp} introduces structural clustering pseudo-labeling in heterogeneous graphs. RMR ~\cite{rmr} divides heterogeneous graphs into multiple relational subgraphs and uses heterogeneous neighbors to reconstruct target node features to learn heterogeneity. HERO ~\cite{hero} introduces self-expression matrices to capture homogeneity in heterogeneous graphs. SCHOOL~\cite{school} optimizes the affinity matrix through a spectral clustering that focuses on homophilic information. Though both types of methods mentioned above have achieved promising performance, these methods ignore the problem of heterophily of heterogeneous graphs, leading to confusion in the representation of different classes of nodes.

\subsection{Heterophily in Homogeneous and Heterogeneous Graph}
The heterophily was first proposed on homogeneous graphs. Related study ~\cite{heterophily,heterophily-s}discusses the performance degradation of earlier GNNs on heterophilic graphs, and ~\cite{fagcn,ACM-GCN} integrates low-pass and high-pass filters to achieve adaptive learning on both homophilic and heterophilic graphs. GraphACL ~\cite{graphacl} and NWR-GAE ~\cite{nwrgae} use neighborhood reconstruction from self-supervised learning to mitigate the effect of heterophily. Greet ~\cite{greet} introduces an edge discriminator to distinguish between homophilic and heterophilic edges. All these studies were conducted on homogeneous graphs. Multiple meta-path subgraphs lead to huge computation when generalizing to heterogeneous graphs.

Recently, several studies have been exploring heterophily on heterogeneous graphs. HDHGR~\cite{hdhgr} utilizes a meta-path similarity learner for structure learning in meta-path subgraphs, and LatGRL~\cite{latgrl} transforms heterogeneous graphs into multiplex graphs and utilizes similarity mining methods to learn the semantic heterophily. These methods preprocess heterogeneous graphs into meta-path subgraphs, losing structural heterogeneity (e.g., actor-movie). Hetero2Net ~\cite{hetero2net} decouples homophilic and heterophilic information by reconstructing positive and negative meta-paths, assumes homophilic properties for metapath-connected nodes, but overlooks potential heterophilic pairs. These methods construct homophilic and heterophilic graphs from the homogeneous graph perspective, which relies on training on predefined homogeneous subgraphs and ignores the effect of heterogeneous edge information.

\begin{figure*}[htbp]
\centering   
\includegraphics[width=1.0\linewidth]{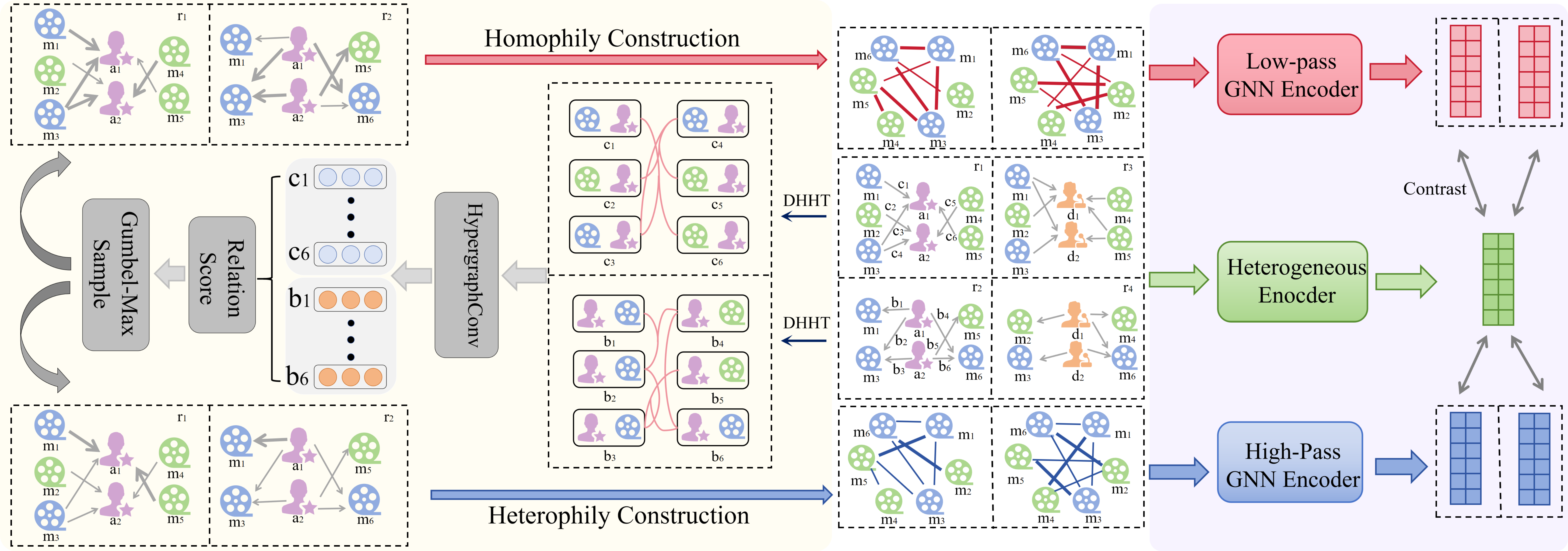} 
\caption{The overall framework of the proposed model (RASH). It utilizes the importance of the heterogeneous edges obtained from the dual heterogeneous hypergraph transform(DHHT) to separate homophilic and heterophilic graphs dynamically. We use \textcolor{gray}{gray} arrows for relationship importance, \textcolor{red}{red} lines for homophilic edges, \textcolor{blue}{blue} lines for heterophilic edges, and different thicknesses for weights. We give an example based on the relationship between movie-actor($r_1$) and actor-movie ($r_2$). Further, contrastive learning guided by different relations injects homophilic and heterophilic information into the heterogeneous representations.} 
\label{Fig.model}
\end{figure*}

\section{Preliminaries}
\label{sec:prelim}

% \begin{definition}
% \textbf{Heterogeneous Graph.}
% A hypergraph graph is defined as: $\mathcal{G} = (\mathcal{V}, \mathcal{E}, \mathcal{A}, \mathcal{R}, \phi, \psi)$, where $\mathcal{V}$ is the set of nodes, $\mathcal{E}$ is the set of edges, $\phi$ : $\mathcal{V} \to \mathcal{A}$ and $\psi$ : $\mathcal{E} \to \mathcal{R}$ are respectively the node type mapping function and the edge type mapping function, $\mathcal{A}$ denotes the set of object types, and $\mathcal{R}$ denotes the set of relations (link types), where $|\mathcal{A}|+|\mathcal{R}|>2$. $\mathbf{A}^R$ is the adjacency matrix under relation $R \in \mathcal{R}$.
% \end{definition}

\begin{definition}
\textbf{Heterogeneous Graph.}
A heterogeneous graph is defined as: $\mathcal{G} = (\mathcal{V}, \mathcal{E}, \mathcal{X}, \mathcal{Z}, \mathcal{A}, \mathcal{R})$, where $\mathcal{V}$ is the set of nodes, $\mathcal{E}$ is the set of edges, $\mathcal{X}$ denotes the set containing all the feature matrices associated with each type of nodes. $\phi$ : $\mathcal{V} \to \mathcal{A}$ and $\psi$ : $\mathcal{E} \to \mathcal{R}$ are respectively the node type mapping function and the edge type mapping function, $\mathcal{A}$ denotes the set of node types, and $\mathcal{R}$ denotes the set of relations (edge types), where $|\mathcal{A}|+|\mathcal{R}|>2$. $\mathcal{X} = \{\mathbf{X}^{1}, ..., \mathbf{X}^{A}, ..., \mathbf{X}^{|\mathcal{A}|}\}$ denotes the features of all types of nodes. $\mathcal{Z} = \{\mathbf{Z}^{1}, ..., \mathbf{Z}^{R}, ..., \mathbf{Z}^{|\mathcal{R}|}\}$ denotes the features of all types of edges.  $\mathbf{A}^R$ is the adjacency matrix under relation $r \in \mathcal{R}$. We can also have $\mathcal{E} = \{\mathbf{A}^{1}, ..., \mathbf{A}^{r}, ..., \mathbf{A}^{|\mathcal{R}|}\}$.
\end{definition}

\section{Methodology}
\label{sec:method}
In this section, we introduce RASH, aiming to separate homophily and heterophily from a heterogeneity perspective. As illustrated in Figure ~\ref{Fig.model}, RASH consists of four main modules: (1) heterogeneous graph encoder to extract node heterogeneity information, (2) relation-aware edge higher-order semantic learning, (3) dynamic homophily-heterophily separation, and (4) multi-relations contrastive learning.

\subsection{Heterogeneous Graph Encoding}
We utilize a heterogeneous graph neural network (HGNN) to extract node representations. Similar to previous work ~\cite{ie-hgcn,shgn}, we divide the representation learning process of the heterogeneous graph into node aggregation and type aggregation. Specifically, in the node aggregation process, the original heterogeneous graph is decomposed into multiple bipartite subgraphs based on the edge type $r$. Each bipartite subgraph contains two different types of nodes. Then, we aggregate each subgraph separately, and for node $i$, the information of different kinds of one-hop neighbors is aggregated based on the edge type $r$ to obtain the relation representation.

\begin{equation}
\mathbf{h}_{i}^{r} = \delta \Big(\mathbf{W}_{s} \mathbf{h}_{i} + \text{Node-Agg} \big(\{\mathbf{W}_{r} \mathbf{h}_{j} | j \in \mathcal{N}_{i}^{r}\}\big) \Big)
\end{equation}
where $\mathcal{N}_{i}^{r} = \{j | (i,j) \in \mathcal{E}, r \in \mathcal{R} \}$ denotes the set of one-hop neighbors of node $i$ under edge type $r$. $W_s, W_r$ are trainable parameters used to project node $i$ and node $j$ of different j types into the same space. \text{Node-Agg} denotes the method of node aggregation and $\delta$ represents the activation function. In the type aggregation process, we fuse the relation representation under multiple bipartite subgraphs corresponding to node $i$ to obtain a heterogeneous node representation.

\begin{equation}
\mathbf{h}_{i}^{\prime} = \text{Type-Agg} \big(\{\mathbf{h}_{i}^{r} | r \in \mathcal{R}, \mathcal{N}_{i}^{r} \neq \varnothing \}\big)
\end{equation}
where \text{Type-Agg} represents the fusion of multiple relation representation. It can be implemented in various ways: mean, sum, concat, and attention. $\mathbf{h}_{i}^{\prime}$ captures the heterogeneity of the heterogeneous graph by aggregating the heterogeneous neighborhood.

\subsection{Relation-Aware Higher-Order Semantic Learning}
The direct splicing of neighboring nodes as edge representations ignores the interaction of heterogeneous edges, and the inability to model higher-order relationships leads to weaker edge-type information, and to capture the relational importance of heterogeneous edges. We first transform a heterogeneous graph into a dual heterogeneous graph and then utilize a hypergraph neural network to extract node representations, i.e., edge representations in the original graph.

\subsubsection{Constructing Dual Heterogeneous Hypergraph}
\begin{definition}
\textbf{Dual Heterogeneous Hypergraph.}
A dual hypergraph ~\cite{dht} is defined as: $\mathcal{G}' = (\mathcal{V}', \mathcal{E}', \mathcal{X}', \mathcal{Z}')$, where $\mathcal{V}$ is the set of nodes, $\mathcal{E}$ is the set of hyperedges. $\mathcal{V}' = \mathcal{E}$.
$\mathcal{X}' = \mathcal{Z}$. $\mathcal{Z}' = \mathcal{X}$.
\end{definition}

A heterogeneous $\mathcal{G}$ can be split into multiple bipartite graphs without losing any structural information. Here, instead of using a bipartite adjacency matrix to describe a bipartite graph, following~\cite{dht}, we define an incidence matrix to describe the structural information contained in the bipartite graph.

Each relation $r \in \mathcal{R}$ corresponds to a bipartite graph contained in the original heterogeneous graph $\mathcal{G}$. Let us assume that relation $r$ involves two node types $A_1$, and $A_2$. That is, the node set $\mathcal{V}^{A_1}$ and the node set $\mathcal{V}^{A_2}$ exist edges $\mathcal{E}^{r}$ between them. We define the incidence matrix as $\mathbf{M}^{r} \in \{0,1\}^{(|\mathcal{V}^{A_1}| + |\mathcal{V}^{A_2}|) \times |\mathcal{E}^{r}|}$. If node $i \in \mathcal{V}^{A_1}$ and node $j \in $ are connected by edge $e \in $ in the original graph, then we have:
\begin{equation}
\mathbf{M}^{r}_{i,e} = 1 \text{ \& } \mathbf{M}^{r}_{j,e} = 1, \text{iff } \mathbf{A}^{r}_{i,j} = 1, <i, j> = e
\end{equation}

% Please see Figure~\ref{} for the illustration of incidence matrix $\mathbf{M}$.
Then, the dual heterogeneous hypergraph can be denoted as follows:

\begin{equation}
\mathcal{E}' = \{\overline{\mathbf{M}}^{1}, ..., \overline{\mathbf{M}}^{r}, ..., \overline{\mathbf{M}}^{|\mathcal{R}|}\}
\end{equation}
where $\overline{\mathbf{M}}^{r}$ denote the transpose matrix of $\mathbf{M}^{r}$

\subsubsection{Heterogeneous Hypergraph Convolution}
For the initial input of the dual heterogeneous hypergraph, we obtain different types of node representations through a heterogeneous graph neural network with the same architecture as the encoder but no shared parameters. Let $\overline{\mathbf{h}}$ denote the output of the encoder and concatenate the node representations based on the edge type $r$ to obtain the initial features.
\begin{equation}
    \mathbf{x}_{ij}^{r} =  \overline{\mathbf{h}}_i^{'}||\overline{\mathbf{h}}_j^{'},\psi{(i,j)} \in r
\end{equation}
Then, we execute hypergraph convolution ~\cite{hypergnn, hypergcn} for the heterogeneous dual hypergraph corresponding to each edge type $r$, using two-stage message passing:$(1)$Node-to-Hyperedge Aggregation: nodes within the hyperedge aggregate to obtain the hyperedge representation; $(2)$Hyperedge-to-Node Aggregation: node-involved hyperedge aggregation updating the node representation. The hypergraph encoder is defined as follows:
\begin{equation}
    \mathbf{H}^{r(l)} = \mathbf{\sigma} \left( \mathbf{D}^{r{-1}} \overline{\mathbf{M}}^r \mathbf{W}^r \mathbf{B}^{r{-1}}  \overline{\mathbf{M}}^{r\top} \mathbf{H^r}^{(l-1)} \mathbf{\Theta}^{r(l)} \right)
\end{equation}
where $\mathbf{\Theta}^{r(l)} \in \mathcal{R}^{d \times d'}$ is the projection matrix of layer $l$. $\mathbf{W}^r$ is the weight matrix of the hyperedge. $\mathbf{D}^{r-1}$ and $\mathbf{B}^{r-1}$ denote the diagonal matrices of the node degree and hyperedge degree, respectively. The $\mathbf{H^r}^{(0)}$ is equal to $\mathbf{X^r}$. The activation function $\mathbf{\sigma}()$ is set to PReLU. Edge importance is captured by performing hypergraph convolution on the dual hypergraph, explicitly modeling the higher-order semantic and contextual information of different types of edges. To sample edges that are highly important in constructing potential homophilic graphs and heterophilic graphs, we utilize a scoring function to get the sampling probability of each edge under edge type $r$.
\begin{equation}
    \mathbf{s}^r_{ij} = \text{Score}(\mathbf{h}^r_{ij}) %+ MLP(e_{ij})%局部和全局
\end{equation}
where $\text{Score()}$ is implemented as a simple linear layer. Due to the nontrivial nature of the Bernoulli distribution, the model is not well-trained to capture the importance of different edges, and we introduce the Gumbel-Max~\cite{discrete, gumbel-softmax} reparameterization method to transform the discrete binary distribution into a continuous distribution of soft weights. For a  bipartite subgraph under edge type $r$, the weight of each edge is defined as follows:
\begin{equation}
 w^r_{ij} = \text{Sigmoid} ((\mathbf{s}^r_{ij} +log \delta -log(1-\delta ))/ \tau)
\end{equation}
where $\delta \sim Uniform(0, 1)$ is the sampled Gumbel random variate. $\tau$ is a temperature constant used to control the degree of sharpening of the weight distribution; the closer $\tau$ is to $0$, the more  $w^r_{ij}$ tends to be binary distribution. Then we separated the heterogeneous graph dynamics into homophilic and heterophilic graphs based on the relational importance scores of the obtained bipartite subgraphs.

\subsection{Dynamic Homophily-Heterophily Separation}
To capture the homophily and heterophily of specific types of nodes, we use the relational importance of bipartite subgraphs to construct the potential homophilic and heterophilic graphs, considering the directionality of bipartite subgraphs, in bipartite graphs $G^r$, if $k$ has a higher importance to node $i$, and in bipartite subgraph $G^{r^{-1}}$, $j$ has a higher importance to $k$. This implies that there are deeper latent connections between $(i,j)$, such as movies of the same genre or papers on the same topic. Therefore, they are more likely to belong to the homophilic node pairs, and on the contrary, it is more likely to belong to the heterophilic node pairs. Based on the relation importance, we compute the weights of the homophilic and heterophilic graphs guided by the edge type $r$:
\begin{equation}
    a^{r,ho}_{i,j} = \sum_{k \in \mathcal{N}_{j}^{r}}w^r_{i,k}w^{r^{-1}}_{k,j}
\end{equation}

\begin{equation}
    a^{r,he}_{i,j} = \sum_{k \in \mathcal{N}_{j}^{r}}(1-w^r_{i,k})(1-w^{r^{-1}}_{k,j})
\end{equation}
where $a^{r,ho}_{i,j}$ and $a^{r,he}_{i,j}$ are the node pair weights of the homophilic and heterophilic graphs consisting of specific types of nodes, and $r^{-1}$ denotes the inverse relationship of r. Unlike latGRL~\cite{latgrl} which needs to pre-process the similarity matrix $topk$ sampling to get the refined multiplex graph before training, we adopt an end-to-end approach to train the potential homophilic and heterophilic graphs, which reduces the dependence on hyperparameters. Also based on relational importance, we can get homogeneous graphs under any type of node without retraining for different types of nodes. This implies that our method provides better scalability.

After the above operations, we obtain multiple homophilic graphs and heterophilic graphs corresponding to the relation $r$. We perform low-pass filtering on the homophilic graphs to preserve the common information between similar nodes, and high-pass filtering on the heterophilic graph to preserve the unique information between different classes of nodes, to avoid feature confusion of the nodes of different classes. We adopt the same encoding method as SGC~\cite{sgc}, the coding process is as follows:
\begin{equation}
    \mathbf{homo}_i^{r,(l)} = \frac{1}{| \mathcal{N}_{i} |} \sum_{j \in \mathcal{N}_{i}}a^{r,ho}_{i,j}x_j^{(l-1)},\mathbf{h}_i^{r,ho} = \mathbf{W}^{r,ho}\mathbf{homo}_i^{r,(l)}
\end{equation}

\begin{equation}
    \mathbf{hete}_i^{r,(l)} = \frac{1}{| \mathcal{N}_{i} |} \sum_{j \in \mathcal{N}_{i}}(x_i-a^{r,he}_{i,j}x_j^{(l-1)}),\mathbf{h}_i^{r,he} = \mathbf{W}^{r,he}\mathbf{hete}_i^{r,(l)}
\end{equation}
where $\mathbf{W}^{r,ho}$ and $\mathbf{W}^{r,he}$ are learned weight
 matrix. $\mathcal{N}_i$ is the set of neighbours of node $i$ in the homogeneous graph. The $\mathbf{h}_i^{r,ho}$ captures the shared information(homophily) between nodes of the homogeneous graph synthesized under the guidance of the heterogeneous relation $r$, while $\mathbf{h}_i^{r,he}$ captures the discrepancy information(heterophily).
% projection layer
%  \begin{equation}
%      \mathbf{z}_i = \text{projection}(\mathbf{h}_i),
%     \mathbf{z}^{r,ho}_{i}= \text{projection}^{r}_{ho}(\mathbf{h}_i^{r,ho}),
%     \mathbf{z}^{r,he}_{i}=\text{projection}^{r}_{he}(\mathbf{h}_i^{r,he})
%  \end{equation}

\subsection{Multi-Relation Contrastive Loss}
Then, we maximize the mutual information of the heterogeneous representations based on the network scheme and the homophilic and heterophilic representations under multiple edge types $r$. Specifically, we regard the heterogeneous representation as anchor points, project the different view representations to the potential space using the learnable MLP, and maximize the consistent information of the other representations using the InfoNCE contrastive loss~\cite{infonce}:
\begin{equation}
    \mathcal{L}(\mathbf{z}_i,\mathbf{z}^{r,ho}_{i}) = -  \frac{1}{|\mathcal{N}_{i}|}\left(\log \frac{\sum_{{v_j} \in \mathcal{P}^{r,ho}_{i}} e^{{\langle}\mathbf{z}_{i}, \mathbf{z}_{j}^{r,ho}\rangle/\tau_{c}}}{\sum_{v_{k} \in \mathcal{V} \setminus \mathcal{P}^{r,ho}_{i}} e^{\langle \mathbf{z}_{i}, \mathbf{z}_{k}^{r,ho}\rangle/\tau_{c}}} \right)
\end{equation}
where $\mathbf{z}_i$ and $\mathbf{z}^{r,ho}_{i}$ denote the projected representation, $\tau_c$ is the temperature constant, $\langle,\rangle $ is the cosine similarity, and $\mathcal{P}^{r,ho}_{i}$ denotes the set of positive samples of node $i$. We utilize the product of the cosine similarity of the obtained homogeneous representation and the edge weights under the corresponding view as the sampling weights: $\mathbf{p}^{r,ho}_{i,j} = a^{r,ho}_{i,j}\langle \mathbf{h}^{r,ho}_{i}, \mathbf{h}^{r,ho}_{j}\rangle $. We choose the top $k$ nodes with the highest similarity to node $i$ as their positive samples. 

To learn the shared information between the nodes in the homophilic view and the discriminative information of the nodes themselves in the heterophilic view, we contrast the anchored representations to learn them separately, and the loss of the complete view under a single relationship $r$ has a composition of homophilic and heterophilic perspectives, expressed as follows:
\begin{equation}
    \mathcal{L}_r^{ho} = \mathcal{L}(\mathbf{z}_i,\mathbf{z}^{r,ho}_{i}) + \mathcal{L}(\mathbf{z}^{r,ho}_{i},\mathbf{z}_i)
\end{equation}
\begin{equation}
    \mathcal{L}_r^{he} = \mathcal{L}(\mathbf{z}_i,\mathbf{z}^{r,he}_{i}) + \mathcal{L}(\mathbf{z}^{r,he}_{i},\mathbf{z}_i)
\end{equation}
The overall loss consists of all the relations $\mathcal{R}$ corresponding under the current node type.
\begin{equation}
    \mathcal{L}_{c} = \sum_{r \in \mathcal{R}}(\mathcal{L}_r^{ho} + \mathcal{L}_r^{he})
\end{equation}
By optimizing the above loss, we use relational importance to bootstrap the synthetic graph to learn the homophilic and heterophilic information under different semantic views while preserving the structural heterogeneity of the heterogeneous graph. In contrast, the method based on multiple meta-path views preserves semantic heterogeneity, but the heterogeneity of nodes is discarded for a single node type. However, the RASH introduces a new approach to learning relational importance through higher-order information of heterogeneous edges, by which we can consider both homophilic and heterophilic information of multiple types of nodes.

\begin{table*}[ht]
\tabcolsep=0.06cm
\begin{center}
\caption{Comparison among all the methods on the node classification task. The highest, and second highest results are highlighted in bold red and blue, respectively}
\label{tab:node-classify}
\begin{tabular}{c|c|c|ccccccccccc}
\toprule
Dataset & Metric & Split & HAN & HGT & DMGI & HDMI & HeCo & HGMAE & MEOW & RMR & HERO & LatGRL & \textbf{RASH} \\
\midrule
\multirow{9}{*}{DBLP} & \multirow{3}{*}{Micro} 
& 20 & 90.16$\pm$0.9 & 89.68$\pm$1.1 & 90.78$\pm$0.3 & 91.44$\pm$0.5 & 91.97$\pm$0.2 & 92.71$\pm$0.5 & 93.06$\pm$0.4 & 92.35$\pm$0.3 & 92.07$\pm$0.2 & \textcolor{blue}{93.73$\pm$0.3} & \textcolor{crimson1}{95.12$\pm$0.1} \\
& & 40 & 89.47$\pm$0.9 & 89.32$\pm$1.2 & 89.92$\pm$0.4 & 90.51$\pm$0.3 & 90.76$\pm$0.3 & 92.43$\pm$0.3 & 91.77$\pm$0.2 & 91.25$\pm$0.4 & 91.14$\pm$0.3 & \textcolor{blue}{92.96$\pm$0.2} & \textcolor{crimson1}{94.04$\pm$0.2} \\
& & 60 & 90.34$\pm$0.8 & 89.75$\pm$0.8 & 90.35$\pm$0.8 & 91.27$\pm$0.3 & 91.59$\pm$0.2 & 93.05$\pm$0.3 & \textcolor{blue}{94.13$\pm$0.2} & 93.27$\pm$0.3 & 91.62$\pm$0.2 & 93.31$\pm$0.3 & \textcolor{crimson1}{95.28$\pm$0.2} \\
\cmidrule{2-14}
& \multirow{3}{*}{Macro} 
& 20 & 89.31$\pm$0.9 & 88.78$\pm$1.3 & 89.94$\pm$0.4 & 90.86$\pm$0.4 & 91.28$\pm$0.2 & 92.28$\pm$0.5 & 92.57$\pm$0.4 & 92.04$\pm$0.3 & 91.68$\pm$0.3 & \textcolor{blue}{93.11$\pm$0.2} & \textcolor{crimson1}{94.82$\pm$0.1} \\
& & 40 & 88.87$\pm$1.0 & 88.60$\pm$1.2 & 89.25$\pm$0.4 & 90.71$\pm$0.5 & 90.34$\pm$0.3 & 92.12$\pm$0.3 & 91.47$\pm$0.2 & 90.93$\pm$0.4 & 91.33$\pm$0.3 & \textcolor{blue}{92.76$\pm$0.1} & \textcolor{crimson1}{93.72$\pm$0.2} \\
& & 60 & 89.20$\pm$0.8 & 88.97$\pm$1.0 & 89.46$\pm$0.6 & 91.39$\pm$0.3 & 90.64$\pm$0.3 & 92.33$\pm$0.3 & \textcolor{blue}{93.49$\pm$0.2} & 92.76$\pm$0.3 & 91.72$\pm$0.2 & 92.61$\pm$0.3 & \textcolor{crimson1}{94.87$\pm$0.1} \\
\cmidrule{2-14}
& \multirow{3}{*}{AUC} 
& 20 & 98.07$\pm$0.6 & 97.65$\pm$0.7 & 97.75$\pm$0.3 & 98.16$\pm$0.2 & 98.32$\pm$0.1 & 98.90$\pm$0.1 & 99.09$\pm$0.1 & 98.44$\pm$0.1 & 98.25$\pm$0.2 & \textcolor{blue}{99.27$\pm$0.0} & \textcolor{crimson1}{99.44$\pm$0.0} \\
& & 40 & 97.48$\pm$0.6 & 97.18$\pm$0.8 & 97.23$\pm$0.2 & 98.01$\pm$0.1 & 98.06$\pm$0.1 & 98.55$\pm$0.1 & 98.81$\pm$0.1 & 97.91$\pm$0.1 & 98.34$\pm$0.1 & \textcolor{blue}{98.87$\pm$0.1} & \textcolor{crimson1}{99.02$\pm$0.1} \\
& & 60 & 97.96$\pm$0.5 & 97.68$\pm$0.5 & 97.72$\pm$0.4 & 98.53$\pm$0.1 & 98.59$\pm$0.1 & 98.89$\pm$0.1 & \textcolor{blue}{99.41$\pm$0.0} & 99.02$\pm$0.1 & 98.44$\pm$0.1 & 99.14$\pm$0.2 & \textcolor{crimson1}{99.50$\pm$0.0} \\
\midrule
\multirow{9}{*}{ACM} & \multirow{3}{*}{Micro} 
& 20 & 85.11$\pm$2.2 & 78.10$\pm$1.8 & 87.60$\pm$0.8 & 89.18$\pm$0.5 & 88.13$\pm$0.8 & 90.24$\pm$0.5 & 91.82$\pm$0.3 & 89.62$\pm$0.5 & 88.71$\pm$0.8 & \textcolor{blue}{93.11$\pm$0.1} & \textcolor{crimson1}{93.22$\pm$0.2} \\
& & 40 & 87.21$\pm$1.2 & 77.12$\pm$1.5 & 86.02$\pm$0.9 & 88.72$\pm$0.6 & 87.45$\pm$0.5 & 90.18$\pm$0.6 & 91.33$\pm$0.3 & 88.25$\pm$0.5 & 88.65$\pm$0.5 & \textcolor{blue}{92.05$\pm$0.4} & \textcolor{crimson1}{92.10$\pm$0.2} \\
& & 60 & 88.10$\pm$1.2 & 79.60$\pm$2.5 & 87.82$\pm$0.5 & 90.40$\pm$0.9 & 88.71$\pm$0.5 & 91.34$\pm$0.4 & 91.99$\pm$0.3 & 87.44$\pm$0.2 & 89.62$\pm$0.4 & \textcolor{blue}{92.44$\pm$0.1} & \textcolor{crimson1}{92.48$\pm$0.1} \\
\cmidrule{2-14}
& \multirow{3}{*}{Macro} 
& 20 & 85.66$\pm$2.1 & 75.06$\pm$1.7 & 87.86$\pm$0.2 & 89.63$\pm$0.6 & 88.56$\pm$0.8 & 90.66$\pm$0.4 & 91.93$\pm$0.3 & 89.55$\pm$0.5 & 88.37$\pm$1.3 & \textcolor{blue}{93.27$\pm$0.1} & \textcolor{crimson1}{93.41$\pm$0.1} \\
& & 40 & 87.47$\pm$1.1 & 75.49$\pm$1.4 & 86.23$\pm$0.8 & 89.03$\pm$0.4 & 87.61$\pm$0.5 & 90.15$\pm$0.6 & 91.35$\pm$0.3 & 88.38$\pm$0.5 & 88.59$\pm$1.1 & \textcolor{blue}{92.20$\pm$0.3} & \textcolor{crimson1}{92.22$\pm$0.2} \\
& & 60 & 88.41$\pm$1.1 & 77.94$\pm$2.2 & 87.97$\pm$0.4 & 90.59$\pm$0.7 & 89.04$\pm$0.5 & 91.59$\pm$0.4 & 92.10$\pm$0.3 & 87.50$\pm$0.1 & 89.32$\pm$0.6 &\textcolor{blue}{ 92.64$\pm$0.1} & \textcolor{crimson1}{92.67$\pm$0.1} \\
\cmidrule{2-14}
& \multirow{3}{*}{AUC}
& 20 & 93.47$\pm$1.5 & 87.49$\pm$1.4 & 96.72$\pm$0.3 & 97.98$\pm$0.4 & 96.49$\pm$0.3 & 97.69$\pm$0.1 & 98.43$\pm$0.2 & 97.05$\pm$0.2 & 96.48$\pm$0.4 & \textcolor{blue}{98.55$\pm$0.0} & \textcolor{crimson1}{98.72$\pm$0.0} \\
& & 40 & 94.84$\pm$0.9 & 88.02$\pm$1.2 & 96.35$\pm$0.3 & 96.07$\pm$0.5 & 96.40$\pm$0.4 & 97.52$\pm$0.1 & 97.94$\pm$0.1 & 96.69$\pm$0.1 & 96.50$\pm$0.3 & \textcolor{blue}{98.24$\pm$0.2} & \textcolor{crimson1}{98.55$\pm$0.0} \\
& & 60 & 94.68$\pm$1.4 & 90.95$\pm$1.6 & 96.79$\pm$0.2 & 97.78$\pm$0.4 & 96.55$\pm$0.3 & 97.87$\pm$0.1 & 98.40$\pm$0.2 & 96.41$\pm$0.2 & 97.03$\pm$0.2 & \textcolor{blue}{98.73$\pm$0.0} & \textcolor{crimson1}{98.85$\pm$0.0} \\
\midrule
\multirow{9}{*}{IMDB} & \multirow{3}{*}{Micro} 
& 20 & 37.08$\pm$0.6 & 41.40$\pm$2.1 & 51.56$\pm$1.5 & 50.64$\pm$1.7 & 42.13$\pm$0.1 & 48.39$\pm$0.6 & 50.86$\pm$0.3 & 50.39$\pm$1.4 & 40.04$\pm$0.1 & \textcolor{blue}{54.09$\pm$0.6} & \textcolor{crimson1}{58.38$\pm$0.8} \\
& & 40 & 43.25$\pm$1.3 & 44.20$\pm$2.0 & 53.32$\pm$1.5 & 51.77$\pm$1.0 & 47.17$\pm$0.2 & 47.52$\pm$1.0 & 51.18$\pm$0.5 & 50.64$\pm$1.3 & 46.85$\pm$0.2 & \textcolor{blue}{57.54$\pm$0.4} & \textcolor{crimson1}{57.58$\pm$0.6} \\
& & 60 & 49.80$\pm$1.0 & 43.70$\pm$1.8 & 56.00$\pm$1.4 & 54.64$\pm$1.2 & 49.15$\pm$0.1 & 50.95$\pm$0.9 & 53.42$\pm$0.8 & 54.14$\pm$1.0 & 43.79$\pm$0.4 & \textcolor{blue}{59.71$\pm$0.1} & \textcolor{crimson1}{62.04$\pm$0.4} \\
\cmidrule{2-14}
& \multirow{3}{*}{Macro} 
& 20 & 27.50$\pm$1.5 & 40.89$\pm$2.3 & 50.99$\pm$1.5 & 50.33$\pm$1.8 & 36.38$\pm$0.2 & 46.55$\pm$0.4 & 49.19$\pm$0.3 & 48.57$\pm$1.5 & 36.13$\pm$0.1 & \textcolor{blue}{53.21$\pm$0.4} & \textcolor{crimson1}{58.14$\pm$0.8} \\
& & 40 & 37.65$\pm$1.9 & 44.23$\pm$1.8 & 53.26$\pm$1.5 & 50.97$\pm$1.1 & 44.66$\pm$0.3 & 45.86$\pm$1.0 & 49.78$\pm$0.5 & 49.06$\pm$1.4 & 45.64$\pm$0.3 & \textcolor{blue}{56.97$\pm$0.3} & \textcolor{crimson1}{57.20$\pm$0.5} \\
& & 60 & 46.87$\pm$1.6 & 43.61$\pm$1.4 & 55.54$\pm$1.5 & 54.61$\pm$1.3 & 45.87$\pm$0.2 & 48.02$\pm$1.1 & 51.99$\pm$1.0 & 52.93$\pm$0.9 & 42.17$\pm$0.2 & \textcolor{blue}{58.91$\pm$0.1} & \textcolor{crimson1}{61.05$\pm$0.5} \\
\cmidrule{2-14}
& \multirow{3}{*}{AUC} 
& 20 & 62.38$\pm$0.4 & 58.61$\pm$1.5 & 69.85$\pm$1.4 & 69.11$\pm$0.0 & 66.50$\pm$0.0 & 66.70$\pm$0.8 & 71.44$\pm$0.1 & 69.37$\pm$0.7 & 56.73$\pm$0.0 & \textcolor{blue}{75.70$\pm$0.1} & \textcolor{crimson1}{76.97$\pm$0.6} \\
& & 40 & 66.67$\pm$0.5 & 59.93$\pm$1.2 & 70.51$\pm$1.5 & 69.88$\pm$0.0 & 66.75$\pm$0.0 & 67.22$\pm$0.6 & 70.24$\pm$0.3 & 68.69$\pm$1.1 & 63.92$\pm$0.1 & \textcolor{blue}{76.54$\pm$0.2} & \textcolor{crimson1}{76.73$\pm$0.1} \\
& & 60 & 70.15$\pm$0.2 & 60.48$\pm$1.0 & 72.84$\pm$1.1 & 72.30$\pm$0.0 & 70.48$\pm$0.0 & 68.59$\pm$0.6 & 71.58$\pm$0.2 & 71.53$\pm$0.7 & 62.13$\pm$0.7 & \textcolor{blue}{77.93$\pm$0.2} & \textcolor{crimson1}{79.52$\pm$0.2} \\
\midrule
\multirow{9}{*}{YELP} 
& \multirow{3}{*}{Micro} 
& 20 & 88.12$\pm$1.1 & 89.74$\pm$0.3 & 73.19$\pm$1.6 & 75.12$\pm$1.5 & 71.71$\pm$0.5 & 68.45$\pm$3.5 & 66.93$\pm$2.2 & 76.06$\pm$2.2 & 90.90$\pm$0.3 & \textcolor{blue}{92.53$\pm$0.3} & \textcolor{crimson1}{94.00$\pm$0.1} \\
& & 40 & 87.57$\pm$0.3 & 90.97$\pm$0.1 & 74.37$\pm$1.9 & 72.98$\pm$1.0 & 71.43$\pm$1.6 & 66.82$\pm$4.3 & 68.16$\pm$1.5 & 78.89$\pm$1.0 & 92.74$\pm$0.2 & \textcolor{blue}{92.50$\pm$0.4} & \textcolor{crimson1}{93.83$\pm$0.1} \\
& & 60 & 84.53$\pm$2.9 & 88.49$\pm$0.4 & 74.91$\pm$1.2 & 73.25$\pm$0.6 & 72.09$\pm$0.4 & 67.23$\pm$3.4 & 66.56$\pm$2.9 & 78.80$\pm$0.6 & 89.83$\pm$0.4 & \textcolor{blue}{91.46$\pm$0.3} & \textcolor{crimson1}{92.99$\pm$0.5} \\
\cmidrule{2-14}
& \multirow{3}{*}{Macro} 
& 20 & 88.34$\pm$1.4 & 90.24$\pm$0.3 & 68.44$\pm$1.8 & 69.85$\pm$1.6 & 68.70$\pm$0.5 & 60.04$\pm$0.9 & 61.77$\pm$0.4 & 73.18$\pm$1.9 & 91.72$\pm$0.3 & \textcolor{blue}{93.36$\pm$0.2} & \textcolor{crimson1}{94.50$\pm$0.1} \\
& & 40 & 87.82$\pm$0.3 & 91.76$\pm$0.2 & 71.87$\pm$1.5 & 69.31$\pm$0.2 & 68.50$\pm$1.4 & 61.75$\pm$1.4 & 64.88$\pm$0.3 & 77.86$\pm$0.7 & 93.54$\pm$0.3 & \textcolor{blue}{93.44$\pm$0.2} & \textcolor{crimson1}{94.63$\pm$0.1} \\
& & 60 & 84.14$\pm$3.9 & 89.25$\pm$0.3 & 71.43$\pm$1.2 & 68.97$\pm$0.1 & 68.71$\pm$0.2 & 60.99$\pm$0.9 & 60.79$\pm$0.6 & 76.83$\pm$0.5 & 90.53$\pm$0.4 & \textcolor{blue}{92.49$\pm$0.2} & \textcolor{crimson1}{94.17$\pm$0.6} \\
\cmidrule{2-14}
& \multirow{3}{*}{AUC} 
& 20 & 96.51$\pm$0.8 & 97.78$\pm$0.1 & 84.47$\pm$5.1 & 89.22$\pm$0.0 & 89.40$\pm$2.6 & 80.34$\pm$4.1 & 84.63$\pm$0.7 & 91.13$\pm$0.7 & 98.24$\pm$0.1 & \textcolor{blue}{98.44$\pm$0.1} & \textcolor{crimson1}{98.86$\pm$0.0} \\
& & 40 & 96.70$\pm$0.3 & 97.89$\pm$0.1 & 85.20$\pm$5.7 & 91.62$\pm$0.0 & 88.90$\pm$3.1 & 83.15$\pm$3.9 & 86.22$\pm$1.4 & 92.00$\pm$0.4 & 98.35$\pm$0.1 & \textcolor{blue}{98.74$\pm$0.1} & \textcolor{crimson1}{98.96$\pm$0.0} \\
& & 60 & 95.38$\pm$0.5 & 96.30$\pm$0.1 & 88.22$\pm$2.1 & 86.60$\pm$0.0 & 87.06$\pm$2.7 & 83.47$\pm$3.3 & 83.52$\pm$2.7 & 91.93$\pm$0.3 & 97.76$\pm$0.1 & \textcolor{blue}{98.12$\pm$0.1} & \textcolor{crimson1}{98.40$\pm$0.1} \\

\bottomrule
\end{tabular}
\end{center}
\end{table*}

\section{Experiment}
In this section, we conduct comprehensive experiments on four real-world datasets in various downstream tasks to verify the effectiveness of our proposed RASH model.
\subsection{Datasets}
To validate the model's performance, we use four public benchmark datasets: DBLP, ACM, IMDB, and YELP. DBLP~\cite{dblp} and ACM~\cite{magnn} are academic paper heterogeneous graph datasets, IMDB~\cite{magnn} is from the movie dataset, and YELP~\cite{yelp} is a business heterogeneous graph dataset. More details about the datasets are found in Appendix ~\ref{dataset}.

\subsection{Baselines}
We compare the performance of the proposed method with various heterogeneous graph learning methods on different datasets, which can be categorized into two types: supervised methods including HAN~\cite{han}, HGT~\cite{hgt}, and self-supervised learning methods including DMGI~\cite{dmgi}, HDMI~\cite{hdmi}, HeCo~\cite{heco}, HGMAE~\cite{hgmae}, MEOW~\cite{meow}, RMR~\cite{rmr}, HERO~\cite{hero} and LatGRL~\cite{latgrl}.
\subsection{Settings}
To ensure statistically reliable results, all experiments are conducted with 10 independent trials followed by the computation of mean performance metrics and their standard deviations. Following the previous unsupervised representation learning method ~\cite{heco}, the model is trained in a self-supervised manner. Then, we freeze the node representations and train a logistic regression classifier. For dataset splits,  we select 20, 40, and 60 nodes in each class as the training set, and 1000 nodes among the remaining nodes as the validation and test set. We conducted experiments in different downstream tasks, including node classification, node clustering, and similarity search.

We use the Pytorch framework to implement our proposed RASH. we use a two-layer heterogeneous encoder and train the model using the Adam ~\cite{adam} optimizer. The learning rate is tuned from 1e-3 to 5e-4 using grid search, the temperature coefficient $\tau_c$ from 0.4 to 0.8 spaced 0.1, and the feature dimensions from \{64, 128, 256, 512\}. The number of low-pass and high-pass filtering layers is searched in \{1, 2, 3, 4, 5\}. The range of the number of positive samples was set from 0 to 5. For gumbel-max $\tau$ was set to 1.0 and $\delta$ was set to 1e-4. We will further tune all baselines' hyperparameters according to the open code and default settings to obtain the best performance. All implementations were executed on an NVIDIA RTX 4090D GPU with 24GB memory.
 
\subsection{Node Classification}
For node classification, we evaluate the model using three standard evaluation metrics including Macro-F1, Micro-F1, and AUC. Table \ref{tab:node-classify} shows that our proposed method achieves the best classification performance on all datasets. For example, on the homophilic DBLP dataset when the label node is 20, our method reaches $95.12\%$ and $94.82\%$ on Micro and Macro, respectively. On the heterophilic IMDB dataset, our method outperforms the current optimal method LatGRL by $4.29\%$ and $4.93\%$ on Micro and Macro, which proves the superiority of the method and demonstrates that our method is able to better capture the heterophily in heterogeneous graphs through the relational importance construction graph.

\subsection{Similarity Search and Node Clustering}
To further validate the generalization of the learned representation to different downstream tasks, we perform similarity search and node clustering tasks on different datasets.
For similarity search, we evaluate $Sim@k$ metrics (k=5,10),
representing the average percentage of top-k nearest neighbors
sharing identical labels in the embedding space. As shown in Table ~\ref{tab:similarity-search}, RASH achieved the optimal results on both $Sim@5$ and $Sim@10$ metrics for all four datasets. For example,
$Sim@5$ reached 93.70\% on the DBLP dataset, significantly higher than the second-place DMGI (91.24\%). This indicates that RASH has a stronger ability to capture the similarity between nodes by separating potential homophily and heterophily, increasing the similarity of nodes of the same class increasing the difference between nodes of different classes, and thus performing well on the similarity search task.

For node clustering, we use commonly used evaluation metrics including normalized mutual information (NMI) and adjusted Rand index (ARI) ~\cite{cluster}. We perform 10 Kmeans clustering for each dataset and take the mean value. The results are shown in Table ~\ref{tab:clustering}. In the node clustering task, RASH still achieves competitive performance. On the heterophilic datasets IMDB and YELP, RASH significantly outperforms the second place, especially on YELP, where RASH's NMI and ARI metrics are improved by 13.74\% and 14.21\%, respectively, over the second-best unsupervised method HERO. This is because compared with HERO, which learns the same matching information of heterogeneous graphs through a self-expression matrix, RASH also captures the heterophilic information of heterogeneous graphs more deeply and learns the differences between nodes. The superior performance in different downstream tasks indicates that its learned representation has strong task-independent representational capabilities.
\begin{table}[t]
\renewcommand{\arraystretch}{1.0} % 控制行间距
\setlength{\tabcolsep}{1pt} % 控制列间距
\small % 设置字体大小为小号
\centering
\caption{Similarity search performance comparison (Sim@5 and Sim@10).}
\label{tab:similarity-search}
\begin{tabular}{c|cc|cc|cc|cc}
\toprule
\multirow{2}{*}{Method} & \multicolumn{2}{c|}{DBLP} & \multicolumn{2}{c|}{ACM} & \multicolumn{2}{c|}{IMDB} & \multicolumn{2}{c}{YELP} \\
\cmidrule{2-9}
& Sim@5 & Sim@10 & Sim@5 & Sim@10 & Sim@5 & Sim@10 & Sim@5 & Sim@10 \\
\midrule
HDMI & 89.58 & 88.80 & 87.86 & 87.36 & 46.96 & 46.87 & 74.08 & 73.23 \\
DMGI & \textcolor{blue}{91.24} & 89.31 & \textcolor{blue}{89.62} & 88.52 & 46.78 & 45.34 & 74.87 & 73.56 \\
HeCo & 89.42 & 88.91 & 88.72 & \textcolor{blue}{88.56} & 42.37 & 40.21 & 76.45 & 76.31 \\
RMR & 89.64 & 89.21 & 86.02 & 85.17 & 46.58 & 45.07 & 76.70 & 75.64 \\
HERO & 90.20 & \textcolor{blue}{89.50} & 88.34 & 87.41 & 44.00 & 42.17 & 77.33 & 75.17 \\
LatGRL & 89.16 & 88.60 & 88.46 & 88.37 & \textcolor{blue}{51.54} & \textcolor{blue}{49.65} & \textcolor{blue}{92.20} & \textcolor{blue}{91.43} \\
\textbf{RASH} & \textcolor{crimson1}{93.70} & \textcolor{crimson1}{93.19} & \textcolor{crimson1}{91.02} & \textcolor{crimson1}{90.00} & \textcolor{crimson1}{52.06} & \textcolor{crimson1}{51.02} & \textcolor{crimson1}{92.62} & \textcolor{crimson1}{92.07} \\
\bottomrule
\end{tabular}
\end{table}

\begin{table}[ht]
\renewcommand{\arraystretch}{1.0} % 控制行间距
\setlength{\tabcolsep}{4.5pt} % 控制列间距
\small % 设置字体大小为小号
\centering
\caption{Node clustering performance comparison}
\label{tab:clustering}
\begin{tabular}{c|cc|cc|cc|cc}
\toprule
\multirow{2}{*}{Method} & \multicolumn{2}{c|}{ACM} & \multicolumn{2}{c|}{DBLP} & \multicolumn{2}{c|}{IMDB} & \multicolumn{2}{c}{YELP} \\
 & NMI & ARI & NMI & ARI & NMI & ARI & NMI & ARI \\
\midrule
HAN & 60.63 & 64.28 & 67.98 & 74.02 & 9.53 & 9.08 & 64.21 & 67.55 \\
HDMI & 38.00 & 32.01 & 70.07 & 75.39 & 8.03 & 8.38 & 37.53 & 42.16 \\
DMGI & 51.66 & 46.64 & 70.06 & 75.46 & \textcolor{blue}{9.41} & \textcolor{blue}{10.12} & 36.95 & 32.56 \\
HeCo & 56.87 & 56.94 & 74.51 & 80.17 & 2.14 & 2.79 & 39.02 & 42.53 \\
HGMAE & \textcolor{blue}{64.57} & \textcolor{blue}{67.85} & \textcolor{blue}{78.47} & \textcolor{blue}{83.09} & 5.55 & 5.86 & 38.95 & 42.60 \\
HERO & 58.41 & 57.89 & 72.52 & 76.75 & 6.95 & 6.16 & \textcolor{blue}{61.19} & \textcolor{blue}{62.87} \\
RMR & 45.39 & 39.29 & 71.72 & 76.95 & 9.22 & 9.22 & 40.70 & 37.02 \\
%LatGRL & 72.52 & 76.75 & 78.79 & 84.28 & 10.30 & 11.90 & 73.71 & 77.40 \\
RASH & \textcolor{crimson1}{68.89} & \textcolor{crimson1}{70.45} & \textcolor{crimson1}{80.28} & \textcolor{crimson1}{85.16} & \textcolor{crimson1}{11.81} & \textcolor{crimson1}{13.02} & \textcolor{crimson1}{74.93} & \textcolor{crimson1}{77.08} \\
\bottomrule
\end{tabular}
\end{table}

\subsection{Ablation Study}
To evaluate the effectiveness of each module. We designed three variants for RASH: w/o Homo\_cl: removing homophilic perspective contrast learning, w/o Hete\_cl: removing heterophilic perspective contrast learning, and w/o RAE, replacing the relational encoding module based on dual heterogeneous hypergraphs with a simple feature concatenation and linear layer. We report the Micro, Macro, and AUC metrics for each class with 20 labeled nodes in Table ~\ref{tab:ablation}. The results show that both homophilic and heterophilic information play important roles in classification, while in the YELP dataset, where heterophilic information dominates, the removal of heterophilic information leads to severe degradation of classification performance. The use of simple edge feature extraction showed a decrease in all metrics, suggesting that dual heterogeneous hypergraph-based relationship encoding is better able to capture higher-order information and interactions of heterogeneous relations.

\begin{table}[t]
\centering
\renewcommand{\arraystretch}{1.0} % 控制行间距
\setlength{\tabcolsep}{2.8pt} % 控制列间距
\small % 设置字体大小为小号
\caption{Ablation study of node classification on the key components of RASH in the different datasets.}
\label{tab:ablation}

\begin{tabular}{c|c|cccc}
\toprule
Metric & Variants & DBLP & ACM & IMDB & YELP \\ \midrule
\multirow{4}{*}{Macro} 
& w/o Homo\_CL & 92.81$\pm$0.1 & 91.65$\pm$0.1 & 49.10$\pm$0.6 & 93.36$\pm$0.0 \\
& w/o Hete\_CL & 92.17$\pm$0.1 & 91.52$\pm$0.2 & 55.58$\pm$0.7 & 85.50$\pm$0.3 \\
& w/o RAE & 93.83$\pm$0.3 & 92.91$\pm$0.2 & 56.63$\pm$0.8 & 92.78$\pm$0.1 \\
& \textbf{RASH} & \textbf{95.12$\pm$0.1} & \textbf{93.41$\pm$0.1} & \textbf{58.14$\pm$0.8} & \textbf{94.00$\pm$0.1} \\ \midrule
\multirow{4}{*}{Micro} 
& w/o Homo\_CL & 93.22$\pm$0.1 & 91.56$\pm$0.1 & 50.69$\pm$0.3 & 92.71$\pm$0.2 \\
& w/o Hete\_CL & 92.59$\pm$0.1 & 91.42$\pm$0.2 & 56.19$\pm$0.7 & 86.29$\pm$0.3 \\
& w/o RAE & 94.20$\pm$0.3 & 92.81$\pm$0.1 & 56.78$\pm$0.7 & 91.94$\pm$0.3 \\
& \textbf{RASH} & \textbf{95.12$\pm$0.1} & \textbf{93.22$\pm$0.2} & \textbf{58.38$\pm$0.8} & \textbf{94.00$\pm$0.1} \\ \midrule
\multirow{4}{*}{AUC} 
& w/o Homo\_CL & 98.71$\pm$0.0 & 98.20$\pm$0.0 & 69.75$\pm$0.3 & 98.61$\pm$0.0 \\
& w/o Hete\_CL & 99.11$\pm$0.0 & 98.34$\pm$0.0 & 75.25$\pm$0.4 & 96.79$\pm$0.0 \\
& w/o RAE & 99.32$\pm$0.0 & 98.67$\pm$0.0 & 75.82$\pm$0.4 & 98.21$\pm$0.1 \\
& \textbf{RASH} & \textbf{99.44$\pm$0.0} & \textbf{98.72$\pm$0.0} & \textbf{76.97$\pm$0.6} & \textbf{98.86$\pm$0.0} \\ \bottomrule
\end{tabular}
\end{table}

\subsection{Effectiveness Analysis}
To demonstrate the effectiveness of our relational encoder based on dual heterogeneous hypergraphs, especially in leveraging higher-order information, we initialize features randomly using 128 dimensional vectors drawn from the Xavier uniform distribution ~\cite{ie-hgcn}. This eliminates meaningful feature information, isolating the impact of higher-order relational structures.

\begin{figure}[ht]
\centering
\subfigure[DBLP]{\includegraphics[width=0.48\columnwidth]{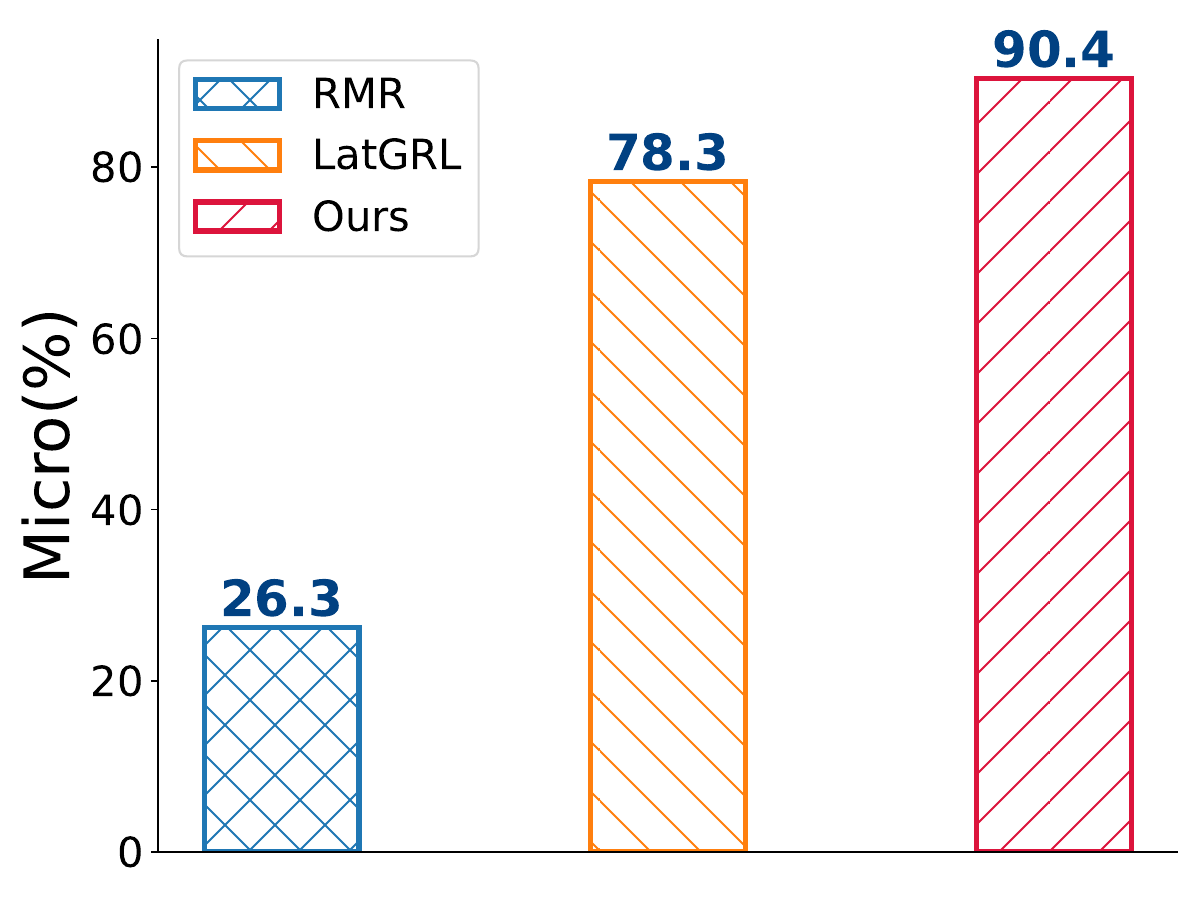}}
\subfigure[IMDB]{\includegraphics[width=0.48\columnwidth]{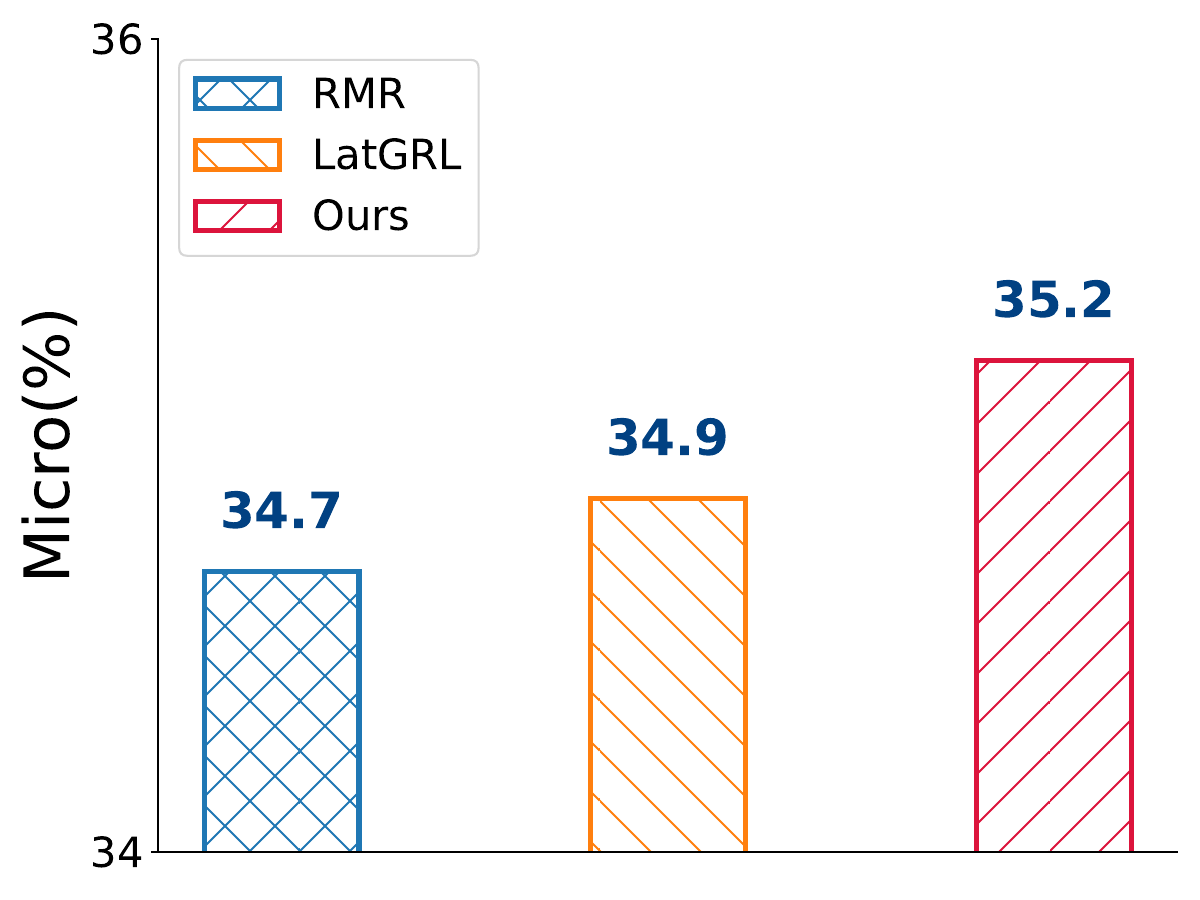}}\\
\subfigure[ACM]{\includegraphics[width=0.48\columnwidth]{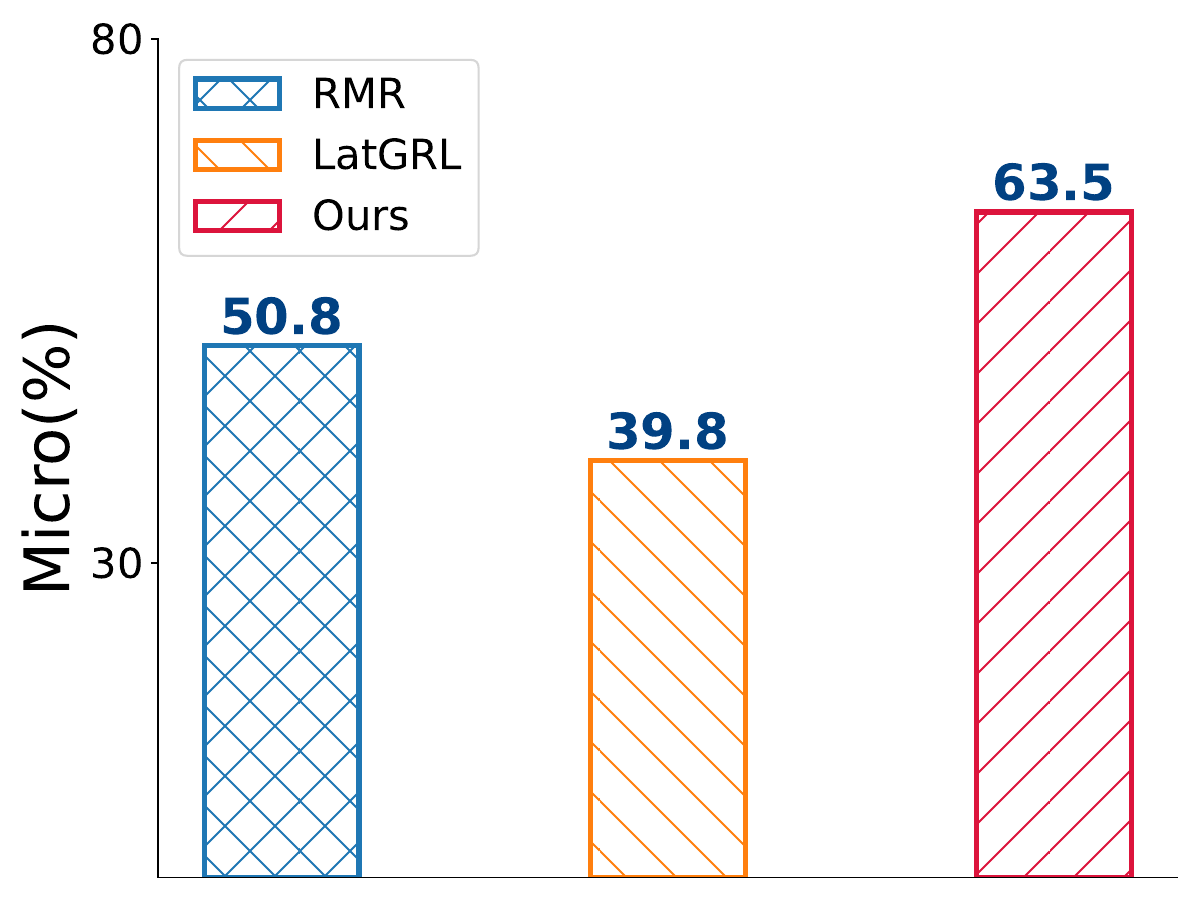}}
\subfigure[YELP]{\includegraphics[width=0.48\columnwidth]{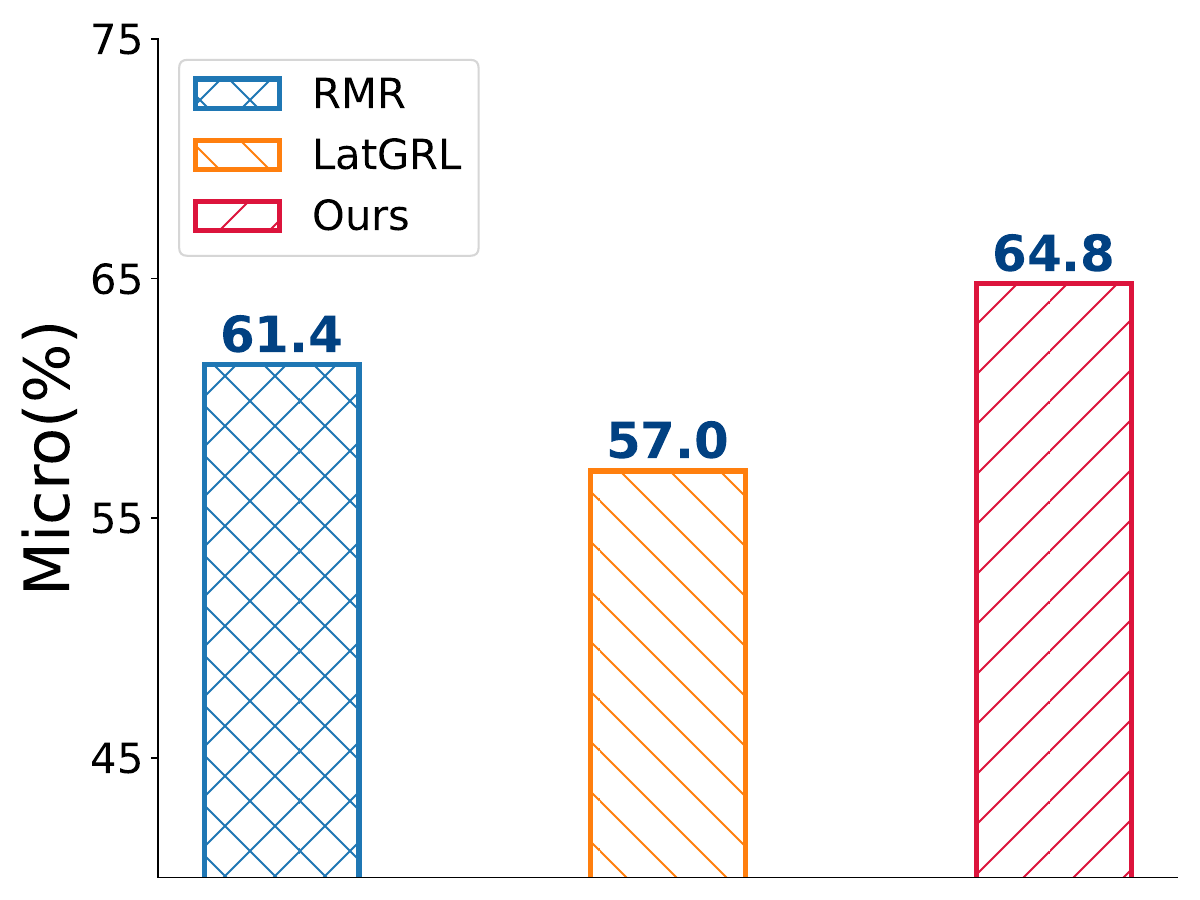}}
\caption{The effect of a randomized feature initialization.}

\label{fig:random_feat}
\end{figure}

We compare our method with state-of-the-art methods and observe changes in Micro metrics. As shown in Figure ~\ref{fig:random_feat}, our method maintains optimal performance even without real features. The RMR based on feature reconstruction and LatGRL based on similarity mining perform poorly without real features, indicating their heavy dependence on original features. In contrast, our method mines potential homophily and heterophily by capturing heterogeneous higher-order structural information, which further proves the effectiveness of our module in capturing meaningful relational patterns even when feature information is limited.

\subsection{Hyper-parameter Analysis}
\begin{figure}[ht]
\centering
\subfigure{\includegraphics[width=0.49\columnwidth]{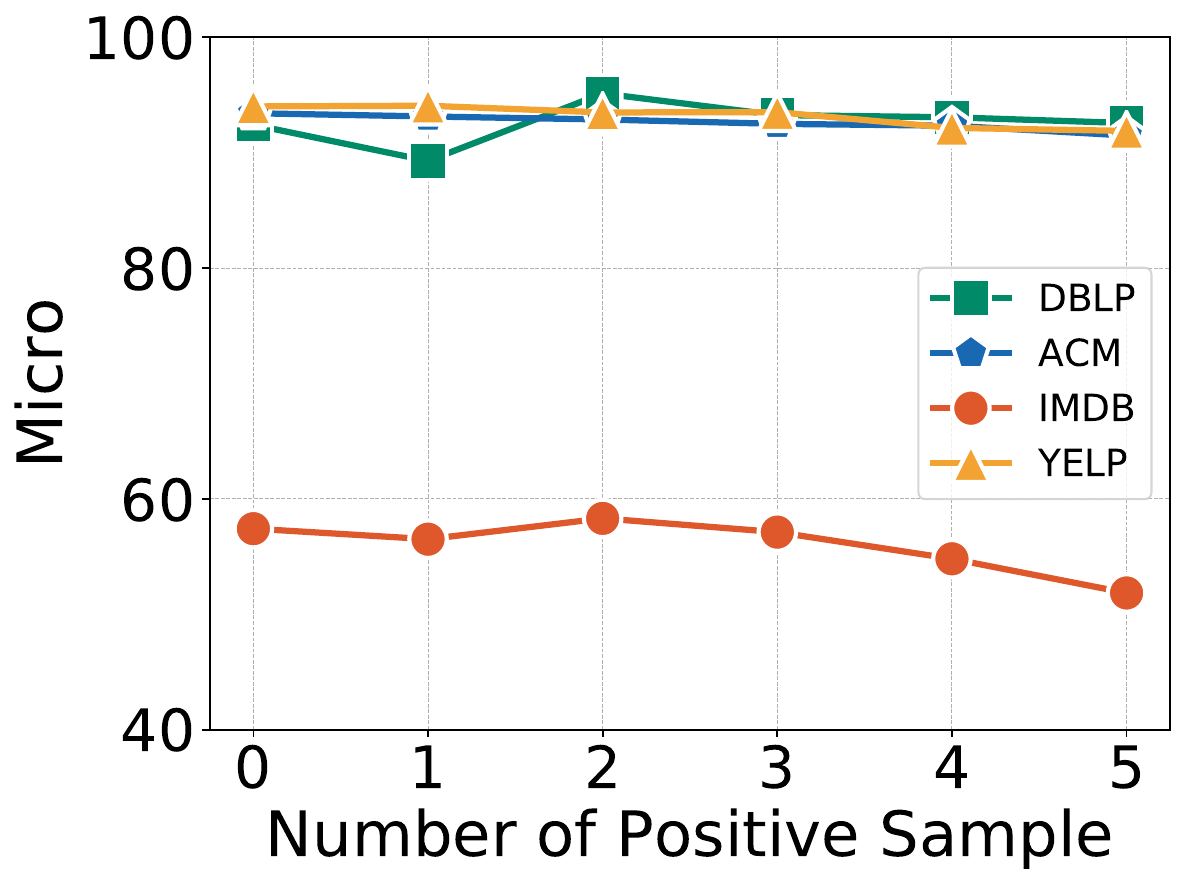}\label{fig:pos}}
\subfigure{\includegraphics[width=0.49\columnwidth]{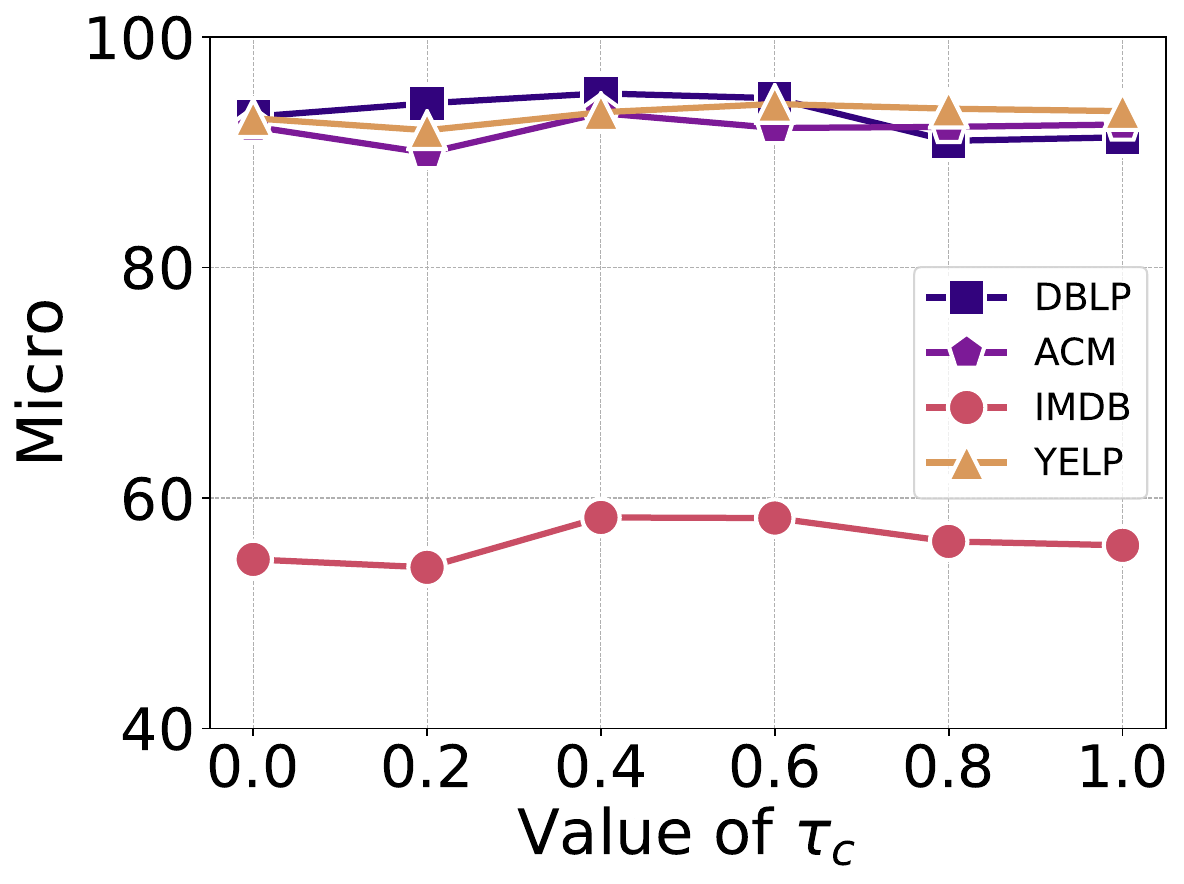}\label{fig:tau}}\\ 
\caption{The sensitivity of hyper-parameter.}
\Description{}
\label{fig:hyper}
\end{figure}

For the hyperparametric analysis, we validate the number of positive samples and temperature constants for contrastive learning, as well as the number of layers in which the constructed homophilic and heterophilic graphs perform low-pass filtering and high-pass filtering. Specifically, we report Micro scores with 20 labeled nodes per class, For the number of samples, we search in the range of [0-5] with a step size of 1. The results are shown in Figure ~\ref{fig:pos}. As the number of samples increases, there is a significant decrease in all four datasets, which suggests that too high several samples may introduce false positive samples. For the temperature constants $\tau_c$, the range is set to [0,1.0], and the step size is 0.2. Figure ~\ref{fig:tau} shows that either too high or too low values affect the classification performance, with ACM, DBLP, and IMDB reaching the optimum at $\tau_c=0.4$, and YELP at $\tau_c=0.6$. 

\begin{figure}[ht]
\centering
\subfigure{\includegraphics[width=0.49\columnwidth]{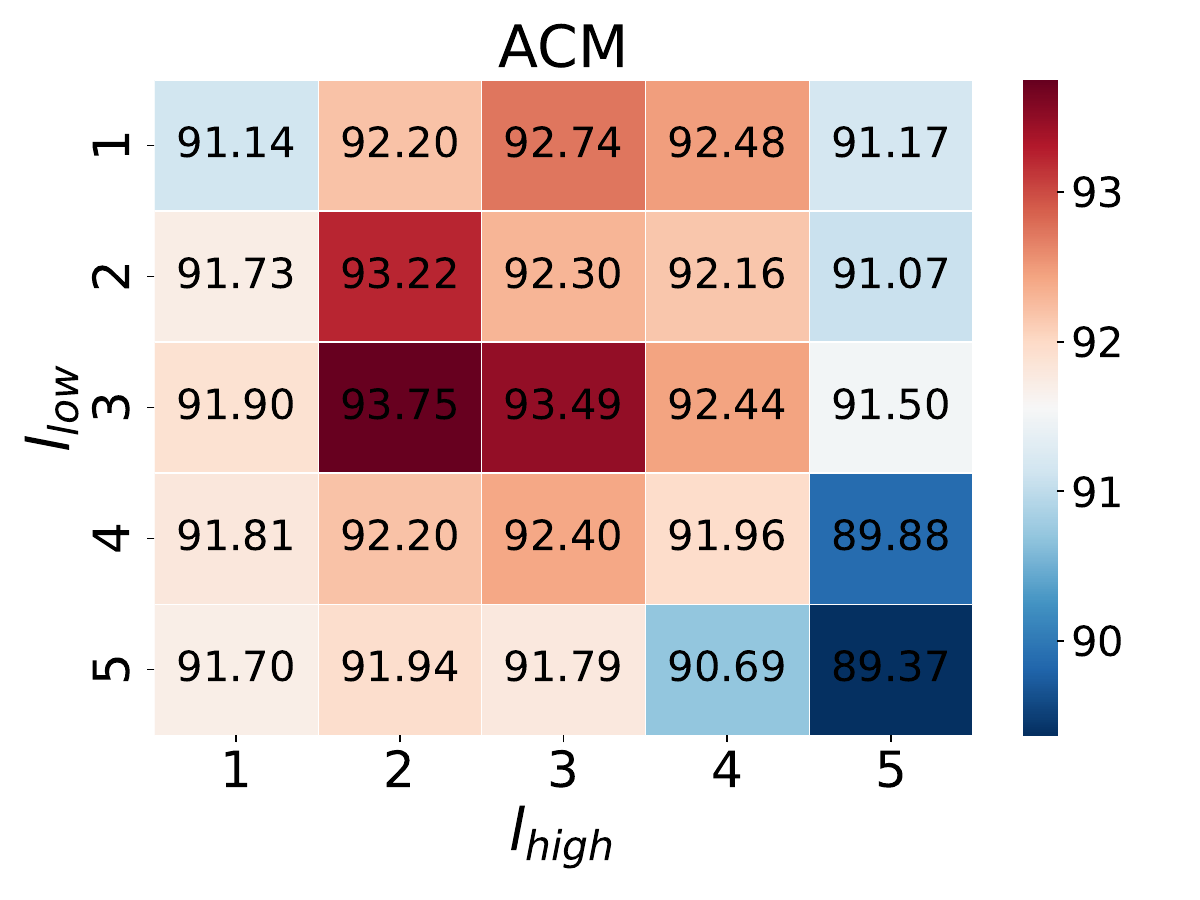}\label{fig:ACM}}
\subfigure{\includegraphics[width=0.49\columnwidth]{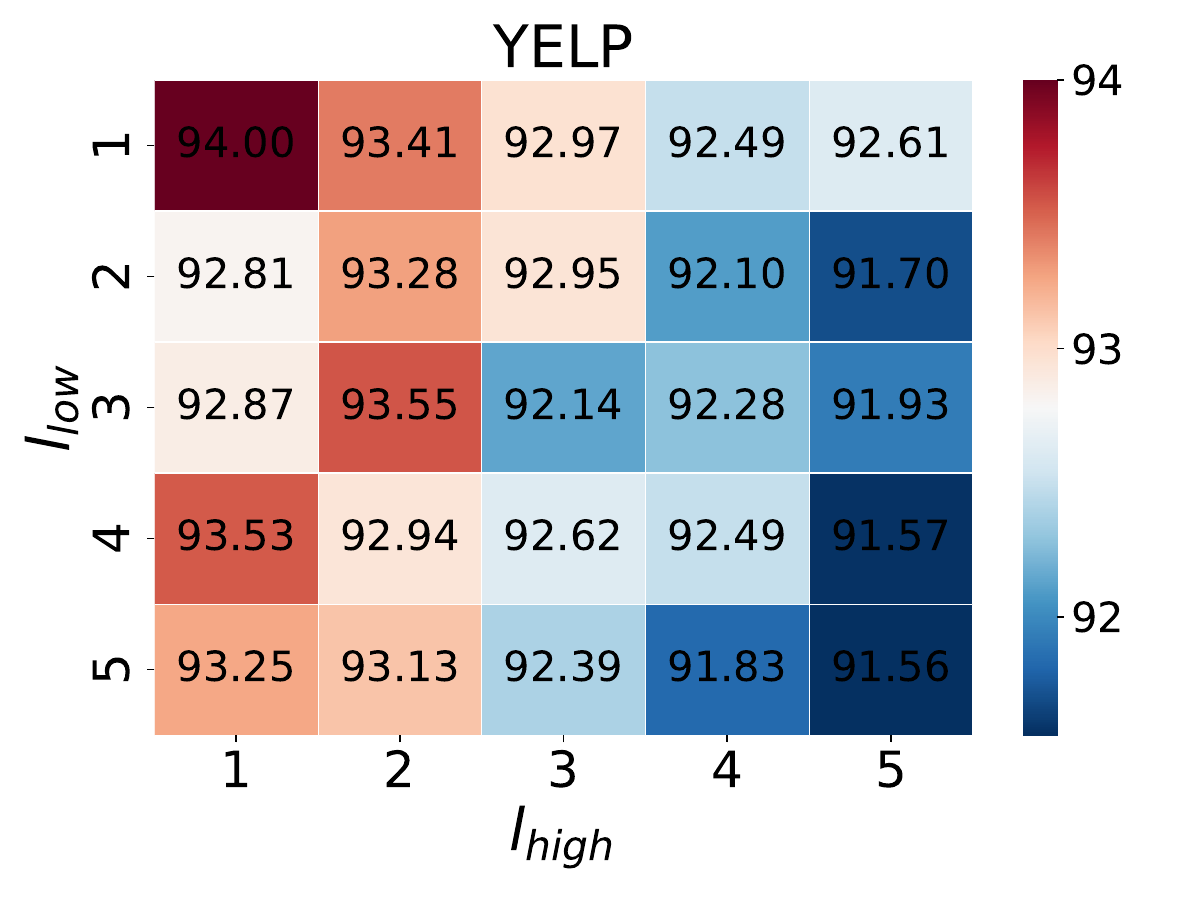}\label{fig:YELP}}\\ 
\caption{The sensitivity of low-pass filtering and
high-pass filtering layers.}
\Description{}
\label{fig:fliter_layer}
\end{figure}

For the effect of the number of layers of low-pass filtering and high-pass filtering, we set the range of the number of layers to [0,5], with a step size of 1, and plotted the heatmap of the two parameters, and the results are shown in Figure ~\ref{fig:fliter_layer}. Whether it's low-pass filtering or high-pass filtering, too high layers lead to performance degradation, and it can be seen that the number of layers in the four datasets all reach the optimum within 2, which avoids high computation caused by too high many filtering layers.

\subsection{Robustness Analysis}
\begin{figure}[ht]
\centering

\subfigure[IMDB]{\includegraphics[width=0.48\columnwidth]{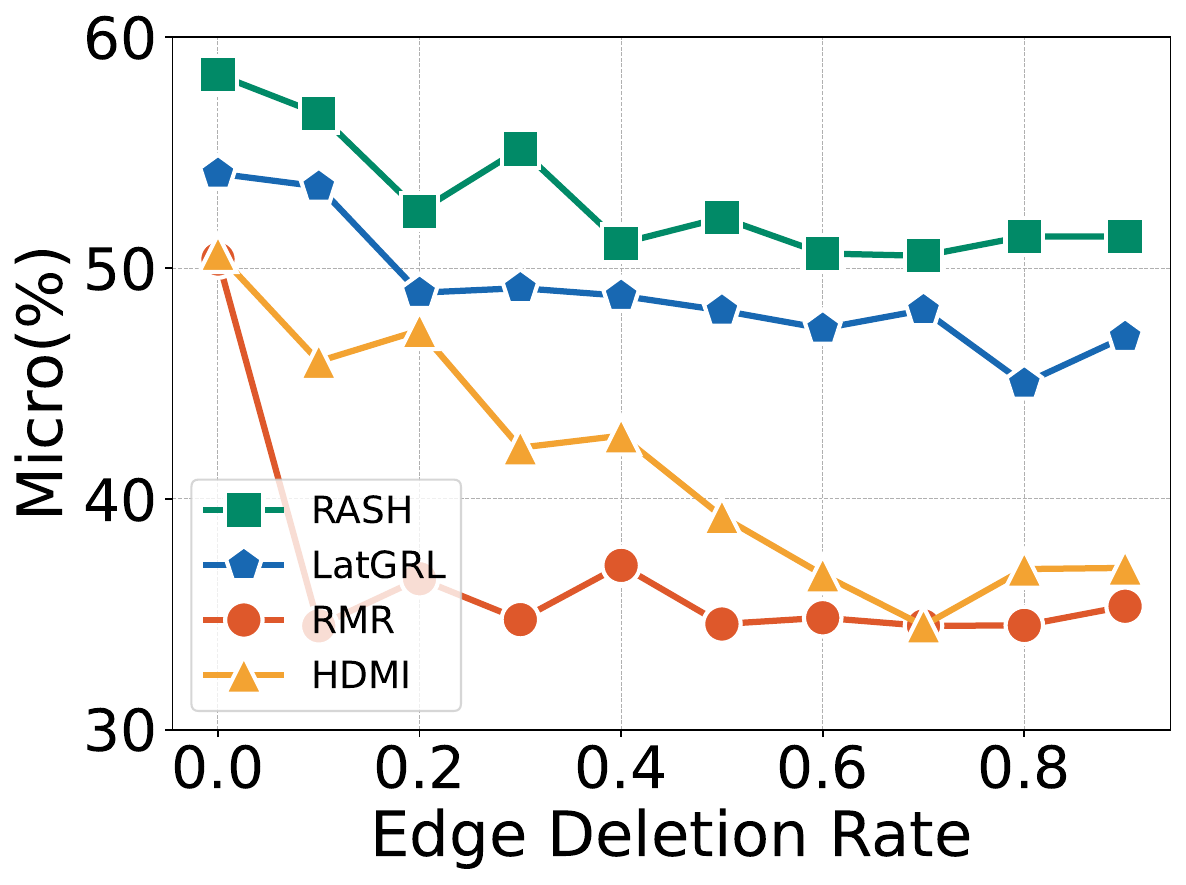}\label{fig:ACM}}
\subfigure[ACM]{\includegraphics[width=0.48\columnwidth]{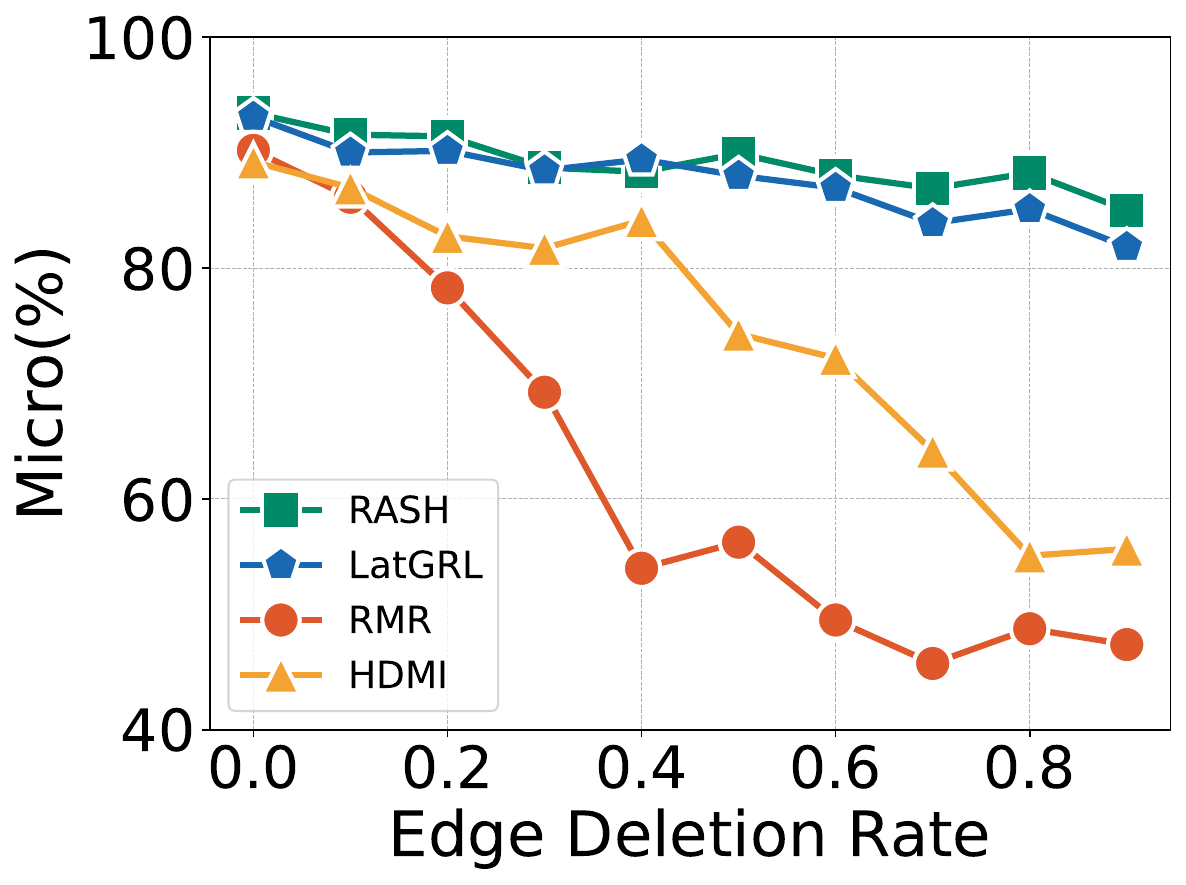}\label{fig:YELP}}
\caption{Robustness analysis on ACM and IMDB}
\label{fig:robust}
\end{figure}

To evaluate the robustness of RASH to topological attacks, we perturb the graph data by randomly removing edges to each bipartite subgraph of the heterogeneous graph. We compare RASH with current state-of-the-art methods LatGRL and RMR to compare the effect of structural noise on the node classification metric Micro when the node training label is 20. We show the performance of node classification in the ACM and IMDB datasets. As can be seen from the Figure~\ref{fig:robust}, as the perturbation rate increases, RASH is relatively stable compared to other methods, especially when the perturbation rate is extremely high, and still performs well. Although our method does not introduce a new structure, it shows better robustness by separating the homophilic and heterophilic information to identify inter-class neighbors and intra-class neighbors.

% \subsection{Complexity Analysis}

% \subsection{Embedding Visualization}

% \subsection{Large Graph}

\section{Conclusion}
In this paper, we investigate the challenge of mining heterophily from a homogeneous perspective with existing heterogeneous graphs without preserving both heterophily and heterogeneity through a real movie actor example. To address this problem, we propose a contrastive learning framework based on the dynamic separation of potential homophily and heterophily in heterogeneous graphs based on relation importance, called RASH, and use the separated homophilic and heterophilic representations for multi-relational contrastive learning to capture homophily and heterophily in heterogeneous graphs. Extensive experiments on four public datasets and different downstream tasks demonstrate the effectiveness of our approach in solving heterogeneous graph heterophilic patterns.

\begin{acks}
This work was supported in part by the National Natural Science Foundation of China under Grants 62425605, 62133012, 62303366, 62472340, and 62203354, in part by the Key Research and Development Program of Shaanxi under Grants 2025CY-YBXM-041, 2024CY2-GJHX-15, 2022ZDLGY01-10, 2025CY-YBXM-042, and 2024GX-YBXM-122, and in part by the Fundamental Research Funds for the Central Universities under Grants ZYTS25086, ZYTS25211, and ZYTS25090.
\end{acks}
%\newpage
%\clearpage

\bibliographystyle{ACM-Reference-Format}
\balance
\bibliography{refs}
%%% -*-BibTeX-*-
%%% Do NOT edit. File created by BibTeX with style
%%% ACM-Reference-Format-Journals [18-Jan-2012].

\clearpage
\appendix
\section{More Details}

\subsection{Datasets}
\label{dataset}
\begin{table}[h!]
\renewcommand{\arraystretch}{1.0} % 控制行间距
\setlength{\tabcolsep}{1pt} % 控制列间距
\centering
\caption{Summary of datasets and their details.}
\begin{tabular}{@{}llcccc@{}}
\toprule
\textbf{Datasets} & \textbf{Node}                & \textbf{Relation} & \textbf{Features} & \textbf{Target} & \textbf{Classes} \\ \midrule
\multirow{4}{*}{DBLP}      
                  & Author (A): 4057             & P-A: 19645        & \multirow{4}{*}{334} & \multirow{4}{*}{author} & \multirow{4}{*}{4} \\
                  & Paper (P): 14328             & P-C: 14328        &                     &                     &                     \\
                  & Conference (C): 20           & P-T: 85810        &                     &                     &                     \\
                  & Term (T): 7723               &                   &                     &                     &                     \\ \midrule
\multirow{3}{*}{ACM}        
                  & Paper (P): 4019              & P-A: 13407        & \multirow{3}{*}{1902} & \multirow{3}{*}{paper}  & \multirow{3}{*}{3} \\
                  & Author (A): 7167             & P-S: 4019         &                     &                     &                     \\
                  & Subject (S): 60              &                   &                     &                     &                     \\ \midrule
\multirow{3}{*}{IMDB}       
                  & Movie (M): 4278              & M-D: 4278         & \multirow{3}{*}{3066} & \multirow{3}{*}{movie}  & \multirow{3}{*}{3} \\
                  & Director (D): 2081           & M-A: 12828        &                     &                     &                     \\
                  & Actor (A): 5257              &                   &                     &                     &                     \\ \midrule
\multirow{4}{*}{YELP}       
                  & Business (B): 2614           & B-U: 30383        & \multirow{4}{*}{82}   & \multirow{4}{*}{business} & \multirow{4}{*}{3} \\
                  & User (U): 1286               & B-S: 2614         &                     &                     &                     \\
                  & Service (S): 4               & B-L: 2614         &                     &                     &                     \\
                  & Rating Levels (L): 9         &                   &                     &                     &                     \\ \midrule
% \multirow{4}{*}{Ogbn-mag}   
%                   & Paper (P): 736389            & P-A: 7145660      & \multirow{4}{*}{128}  & \multirow{4}{*}{paper}   & \multirow{4}{*}{4} \\
%                   & Author (A): 1134649          & P-F: 7505078      &                     &                     &                     \\
%                   & Institution (I): 8740        & A-I: 1043998      &                     &                     &                     \\
%                   & Field (F): 59965             &                   &                     &                     &                     \\ \bottomrule
\end{tabular}
\end{table}
% \section{MORE EXPERIMENTAL RESULTS}
\subsection{Complexity Analysis}
We analyze the time complexity of each module in RASH. Let $r$ denote the number of relations for a specific node type $V_r$ and $E_r$ represent the number of nodes and edges in the corresponding bipartite graph $G^r$ and $b$ denote the batch size for contrastive loss. Due to the sparsity of real-world graphs, both the original heterogeneous graph and hypergraph structures are stored using sparse matrices. 

For the heterogeneous graph encoder: At the l-th layer, let $d$ and $d'$ denote the input and output dimensions of node representations, and $d_p$ denote the projection layer dimension. The time complexity for node-level aggregation is $O(rN_rdd’)$. We adopt summation fusion for type-level aggregation with a time complexity of $O(rN_r)$. For dual heterogeneous hypergraph transformation, which is executed once during preprocessing. Compared to line graphs requiring $O(N_r^2)$, our method achieves $O(rE_r)$ complexity. The message passing complexity depends on the number of edges, resulting in $O(E_r)$ time complexity for hypergraph convolution on each bipartite graph, equivalent to node-level operations on the original graph. Linear layers with small dimensions contribute negligible complexity. The total complexity thus remains $O(rE_r)$. The time complexity is $O(N_r^2)$ for homophily-heterophily separation. The time complexity of multi-relational contrastive loss is $O(N_rbd_p)$.

We compared RASH's efficiency with HGMAE and LatGRL on Yelp. RASH achieves a $1.7 \times$ speedup over HGMAE and uses $31\%$ less memory, highlighting its efficiency. LatGRL has lower time/memory costs, but its efficiency depends on extensive preprocessing (similarity matrix pre-calculation), which isn't included in the reported metrics and limits its applicability in scenarios with limited raw features (Fig.3). In contrast, RASH maintains acceptable efficiency while addressing heterogeneity (structural and semantic) and heterophily.

\begin{table}[h]
\centering
\caption{Time and memory analysis on YELP.}
\begin{tabular}{lccc}
\toprule
 & \textbf{LatGRL} & \textbf{RASH} & \textbf{HGMAE} \\
\midrule
memory(M) & 858 & 1428 & 2071 \\
time(s) & 0.08 & 0.39 & 1.14 \\
\bottomrule
\end{tabular}
\end{table}

% \subsection{Visualization}

\section{More experiments}

\subsection{Effectiveness of Multi-relation Loss}
To verify the role of multi-relational loss. we set up 2 variants: mean view fusion(variant 1) and randomly sampling one view(variant 2). Multi-relation contrast achieves the best performance, proving its necessity, while mean fusion causes semantic blurring, and single-relation fails to capture cross-relation dependencies.

\begin{table}[h]
\centering
\caption{The effect of multi relation contrastive loss.}
\begin{tabular}{l|ccc}
\toprule
  Dataset & \textbf{Variant 1} & \textbf{Variant 2} & \textbf{Multi Relation} \\
\midrule
ACM & 92.30±0.2 & 89.18±0.1 & \textbf{93.22±0.2} \\
IMDB & 55.22±0.5 & 48.20±0.2 & \textbf{58.38±0.8} \\
YELP & 92.85±0.2 & 91.02±0.4 & \textbf{94.00±0.1} \\
\bottomrule
\end{tabular}
\end{table}

\subsection{The Effect of Different Hypergraph Convolutions}
We compare the effect of different hypergraph convolutions on classification performance, such as HyperGCN\cite{hypergcn}, UniSAGE\cite{unignn}, and UniGAT\cite{unignn}. Table~\ref{tab:hyperconv} indicate that UniGAT outperforms others, as attention mechanisms better capture heterogeneous relation importance. This validates RASH’s flexibility in integrating advanced hypergraph operators.

\begin{table}[h]
\centering
\caption{The effect of different hypergraph convolution modules.}
\label{tab:hyperconv}

\begin{tabular}{l|ccc}
\toprule
 Dataset & \textbf{HyperGCN} & \textbf{UniSAGE} & \textbf{UniGAT} \\
\midrule
ACM & 92.10±0.2 & 91.88±0.2 & \textbf{92.42±0.3} \\
IMDB & 57.58±0.6 & 58.80±0.3 & \textbf{59.09±0.2} \\
YELP & 94.22±0.1 & 93.02±0.1 & \textbf{94.89±0.2} \\
\bottomrule
\end{tabular}
\end{table}

\subsection{Large-scale Graph Experiment}
To evaluate the scalability of our method on the large-scale graph, we added the additional large graph experiment Aminer(439k nodes) from RMR~\cite{rmr}. As shown in the Table \ref{tab:node-large}, we report the micro metrics at labeled nodes of 40 and 60. RASH achieves superior performance across different training splits, outperforming RMR by 6.2\% in Micro under the 60-label setting.

We also provide an analysis of the efficiency on large graphs. We report on the hypergraph convolution process and overall computational efficiency on Table ~\ref{tab:time-large}. The computational efficiency of hypergraph convolution across different relations of hyperedges. The memory cost and per-epoch time are acceptable.
\begin{table}[ht]
\begin{center}
\caption{Node classification on the large-scale graph.}
\label{tab:node-large}
\begin{tabular}{c|c|c|cc}
\toprule
Dataset & Metric & Split & RMR  & \textbf{RASH} \\
\midrule
\multirow{4}{*}{Aminer} & \multirow{2}{*}{Micro} 
& 40 & 89.47$\pm$0.9  & \textcolor{crimson1}{94.04$\pm$0.2} \\
& & 60 & 90.34$\pm$0.8 & \textcolor{crimson1}{95.28$\pm$0.2} \\
\cmidrule{2-5}
& \multirow{2}{*}{Macro} 
& 40 & 88.87$\pm$1.0 & \textcolor{crimson1}{93.72$\pm$0.2} \\
& & 60 & 89.20$\pm$0.8 & \textcolor{crimson1}{94.87$\pm$0.1} \\
\bottomrule
\end{tabular}
\end{center}
\end{table}

\begin{table}[h]
\centering
\caption{Time and memory analysis on the large-scale graph.}
\label{tab:time-large}
\begin{tabular}{lcccc}
\toprule
 & \textbf{P-A} & \textbf{P-R} & \textbf{P-C} & \textbf{ALL} \\
\midrule
memory(M) & 380 & 420 & 136 & 9194 \\
time(s) & 0.0016 & 0.0018 & 0.0016 & 0.4173\\
\bottomrule
\end{tabular}
\end{table}

\subsection{Stability Analysis}
\begin{figure}[ht]
\centering

\subfigure[YELP]{\includegraphics[width=0.48\columnwidth]{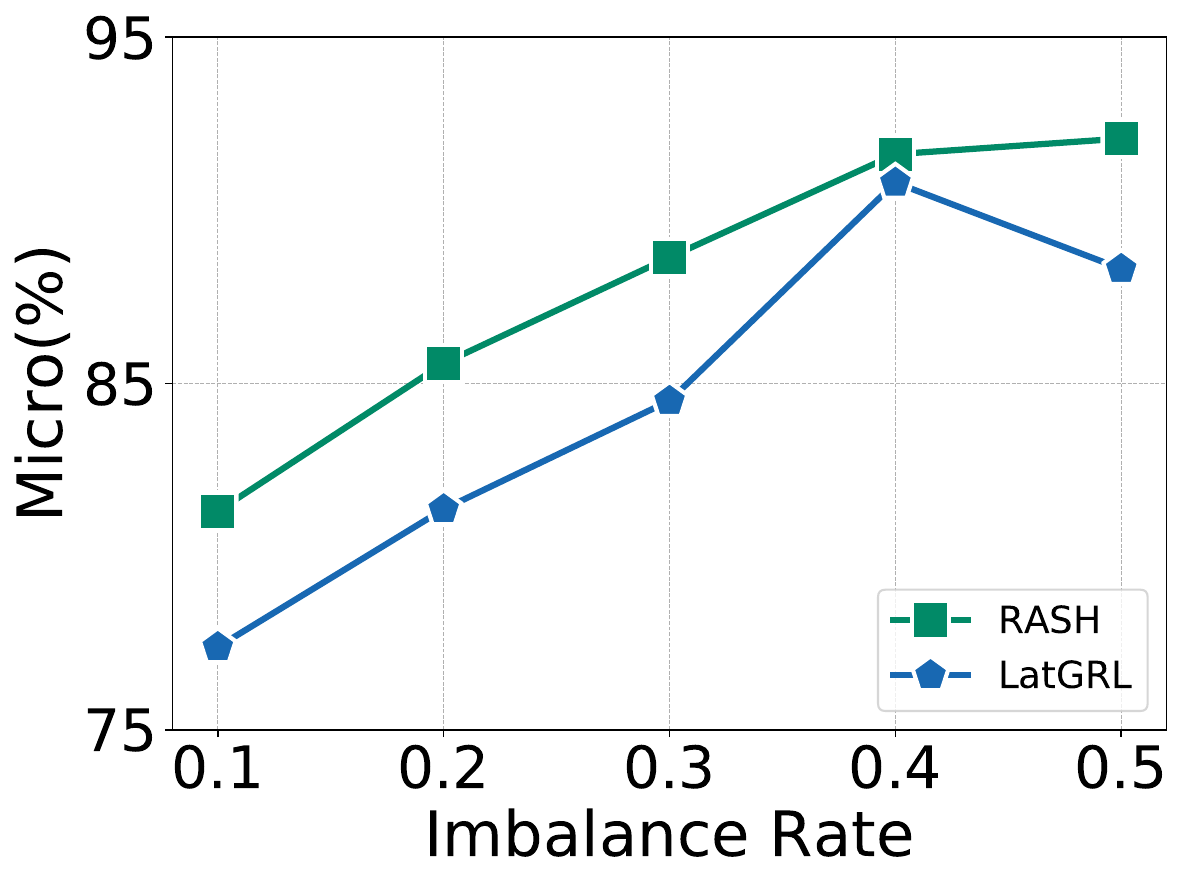}}
\subfigure[ACM]{\includegraphics[width=0.48\columnwidth]{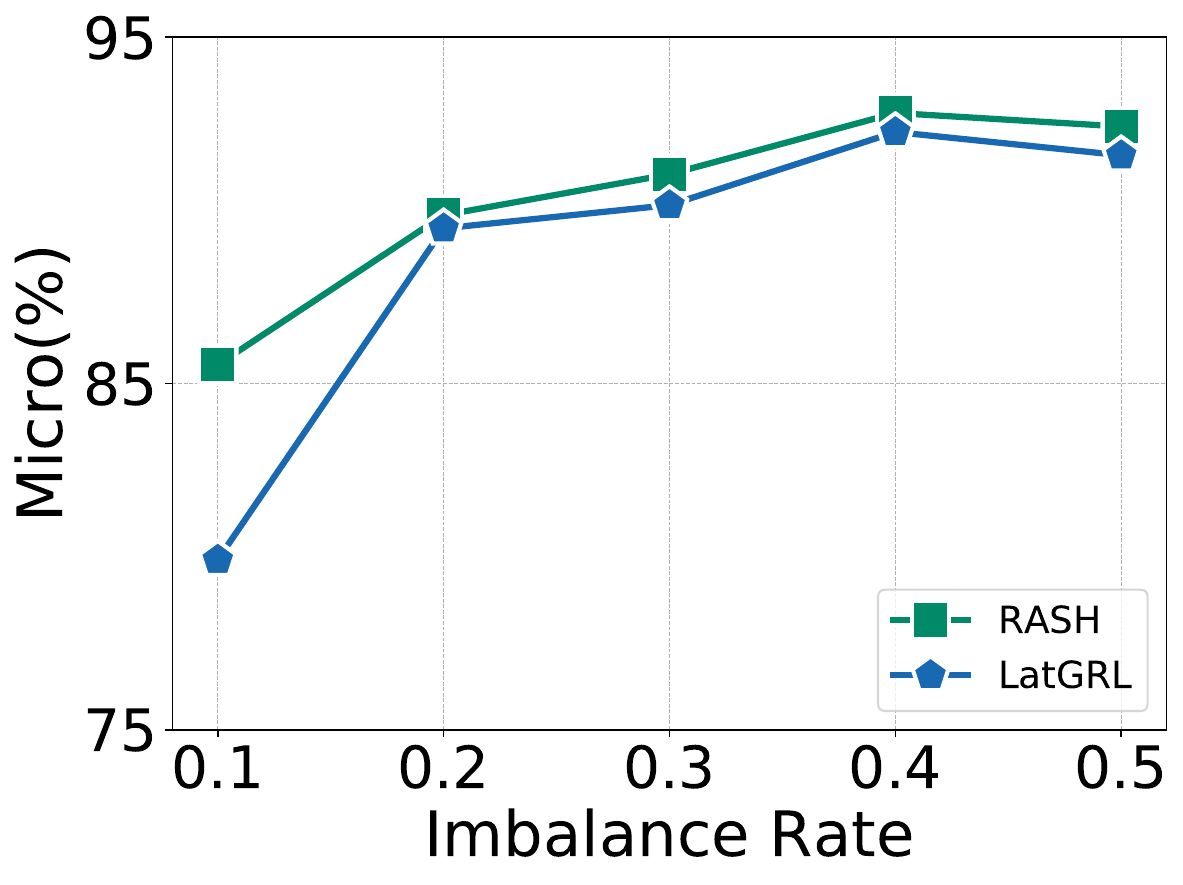}}
\caption{Stability analysis on YELP and ACM.}
\label{fig:stability}
\end{figure}

To evaluate the stability of RASH during data imbalance, we use the same setting as earlier~\cite{graphens} with a labeling rate of $6\%$, and we compare the performance of the imbalance rate at different values, the results are shown in Figure ~\ref{fig:stability}, our method has higher stability in classification performance when the data are imbalanced and exhibits higher performance and stability at extreme imbalance (rate=0.1). The dynamically constructed homophilic and heterophilic maps based on heterogeneous relations provide potential class-related information. In contrast, LatGRL constructs fixed graph structures at the preprocessing stage.

\vfill
\clearpage

\end{document}